\documentclass[11pt]{article}
\usepackage[margin=0.95in]{geometry}
\usepackage{siunitx}
\usepackage{helvet}
\usepackage{amssymb}

\usepackage{graphicx}
\usepackage{subcaption}
\usepackage{url}
\usepackage{amsmath}
\usepackage{tikz}
\usetikzlibrary{positioning, arrows.meta, shapes.geometric, fit, backgrounds, calc}
\usepackage{setspace}

\newenvironment{myprotocol}[1][htb]{%
    \floatname{algorithm}{Protocol}
   \begin{algorithm}[#1]%
   \footnotesize{}
  }{\end{algorithm}}

\usepackage[square,numbers]{natbib}
\usepackage[font=footnotesize]{caption}

\usepackage{parskip}       
\usepackage{titlesec}

\usepackage{soul}
\usepackage{color}
\usepackage{booktabs,multicol,multirow,subcaption}
\usepackage{threeparttable}
\usepackage{tabularx}
\usepackage{algorithm}
\usepackage{algpseudocode}

\newcommand{\piBIN}{\pi_{\text{BIN}}}
\newcommand{\piMARG}{\pi_{\text{MARG}}}
\newcommand{\piPGM}{\pi_{\text{PRIVATE-PGM}}}
\newcommand{\piGAUSS}{\pi_{\text{BATCH-GAUSS}}}

\begin{document}
\setstretch{0.9}  
\setlength{\parskip}{2pt}  
\setlength{\parindent}{1em} 
\setlength{\floatsep}{4pt}       
\setlength{\textfloatsep}{4pt}   
\setlength{\intextsep}{4pt}      

\begin{center}
\large{\textbf{Federated Generation of Synthetic RNA-seq Data}}\\
\vskip 5pt
Daniil Filienko,
Martine De Cock,
Sikha Pentyala\\

\vskip 5pt
{University of Washington Tacoma}
\vskip 5pt
Corresponding author: Daniil Filienko, daniilf@uw.edu

\end{center}
\newpage
\begin{abstract}
    Access to genomic data is highly regulated due to its sensitive nature. While safeguards are essential, cumbersome data access processes pose a significant barrier to the development of AI methods for genomics. Synthetic data generation can mitigate this tension by enabling broader data sharing without exposing sensitive information. Synthetic genomic data are produced by training generative models on real data and subsequently sampling artificial data that preserves relevant statistics while limiting disclosures about the underlying individuals. In some settings, a single data holder may have sufficient data to train such generative models; however, in many applications  data must be combined across multiple sites to achieve adequate scale. This need arises, e.g., in rare disease studies, where individual hospitals typically hold data for only a small number of patients.
    The solution we present in this paper enables multiple data holders to jointly train a synthetic data generator without revealing their raw data. Our approach combines secure multiparty computation (MPC) to ensure input privacy, so that no party ever discloses its data in unencrypted form, with differential privacy (DP) to provide output privacy by mitigating information leakage from the released synthetic data. We empirically demonstrate the effectiveness of the proposed method by generating high‑utility synthetic datasets from multiple real RNA‑seq cohorts in federated settings, showing that our approach enables privacy‑preserving data synthesis even when data are distributed across institutions.

\end{abstract}

\vspace{-10pt}
\section{Introduction}
Patient data, and genomic data in particular, is subject to strict access controls. Even when datasets are advertised as ``open'', researchers are often required to navigate long and cumbersome access procedures, such as negotiating complex contractual agreements with data controllers, completing extensive training requirements, or performing analyses within controlled data enclaves \cite{watson2023delivering}. In contrast, AI researchers are used to minimal barriers when accessing, exploring, and experimenting with new datasets. This mismatch significantly constrains AI work in biomedicine.
At the same time, datasets containing protected health information cannot be made publicly available without careful safeguards. Concerns regarding the privacy and confidentiality of clinical and genomic data are well justified and have led to increasingly strict controlled‑access requirements \cite{NIH-NIST2024,NIH2025,oestreich2021privacy}.

A promising method for operationalizing sensitive patient data for research while protecting patient privacy is synthetic data generation (SDG). A common approach to SDG is to train a generative model on real data and subsequently use the trained model to create artificial data that preserves important characteristics of the real data without disclosing sensitive information about individuals \cite{hu2023sok}. While in some scenarios a single data holder (e.g., a hospital) may have sufficient data on its own to enable SDG, in many cases data from multiple sites needs to be combined. Such federated settings are common, for instance, with rare diseases, where each clinical site has data for a small patient cohort.

\textit{In this paper we present a method for training privacy-preserving synthetic RNA sequencing (RNA-seq) data generators in  federated settings.} Our approach adopts cryptographic techniques based on secure multiparty computation (MPC) \cite{CDN2015} to support a federated training process in which each data holder encrypts their data on site before sending it as input. MPC servers perform training operations on the input data while it stays encrypted, thus providing \textit{input privacy}. Furthermore, we use differential privacy (DP) \cite{dwork2014algorithmic} to provide individual-level privacy guarantees for the released synthetic data, thus providing \textit{output privacy}, 
{\textcolor{black}{exposing only the aggregate population-level bin means necessary for efficient data reconstruction.} 

The heart of our solution \textcolor{black}{are MPC-protocols to run Private-PGM in a federated setting.} Private-PGM is a marginal-based SDG algorithm that learns a probabilistic graphical model (PGM) from marginal probability distributions estimated on (a discretized version of) the real data, and subsequently samples from the model to generate synthetic data  \cite{mckenna2019graphical}. Private-PGM offers formal DP guarantees by perturbing the estimated marginals with noise. It has been shown to work well for generation of synthetic bulk RNA-seq data in \textit{centralized} settings, i.e., where all the real data resides with one data holder \cite{chen2024towards, menziessynthetic__camda}. Our extension of Private-PGM to federated settings consists of cryptographic protocols for quantile binning and estimating marginal probability distributions over the combined data of the data custodians while it remains encrypted. To this end, we draw inspiration from MPC protocols in the \texttt{CaPS} framework previously proposed by Pentyala et al.~\cite{pentyala2024caps} for the marginal-based SDG algorithms MWEM-PGM and AIM \cite{mckenna2019graphical, mckenna2022aim}. We found, however, that a straightforward implementation of these MPC protocols for Private-PGM does not scale well with the high dimensionality of RNA-seq data. Different from Fu et al., who recently proposed to address the inefficiencies of \texttt{CaPS} by optimizing secure sort-and-count operations \cite{towardsfu2025}, we replace iterative ``per sample per gene'' computations by one dot product over a vector of all samples per gene.
Using a variety of TCGA and leukemia datasets, we demonstrate that our approach for federated synthetic RNA-seq data generation can produce synthetic data in as little as a few minutes, depending on the MPC scheme, with a level of quality that is at par with the centralized setting.

\vspace{-10pt}
\section{Preliminaries}
\subsection{Secure Multiparty Computation for Input Privacy}\label{sec:MPCbasics}

Secure multiparty computation (MPC) is an umbrella term for cryptographic approaches that allow two or more parties to jointly compute a specified output from distributed private information without revealing it to one another \cite{CDN2015}. In this paper, we follow the
``MPC as a service'' paradigm, where data holders delegate
the computations to a set of non-colluding, but otherwise
non-trusted servers. We assume the existence of secure communication channels between the servers (parties), as is standard in the MPC threat model.
We adopt secret-sharing based MPC  where each data holder encrypts their private data $x$ by splitting it into shares (see below for an example) and distributing the shares among a set of MPC servers $\mathcal{S} = \{S_1, S_2, \dots, S_K\}$. Though $x$ can be reconstructed when all shares are combined, none of the MPC servers by themselves learn anything about the value of $x$. 
These servers run MPC protocols to jointly perform computations over the secret shares, in our case computations for generating synthetic data. As all computations are done over the secret shares, the servers do not learn the values of the inputs nor of intermediate results, i.e., MPC provides \textit{input privacy}. 

The security guarantees of an MPC protocol are defined by the type of adversary it can withstand. These are typically categorized into passive and active threat models, reflecting the  power an adversary has over the MPC servers it corrupts. 
In the \textit{passive} (or ``semi-honest'') setting, all parties follow the instructions of the protocol, but the adversary attempts to learn private information from the internal state of the corrupted servers and the messages that they receive. MPC protocols that are secure against passive adversaries prevent such leakage of information. 
In contrast, an active (or “malicious”) adversary may corrupt an MPC server to arbitrarily deviate from the protocol, including colluding with others to breach privacy or corrupt intermediate computations. Protection against such adversaries is achieved using information-theoretic message authentication codes (MACs). In addition to holding a share of each secret value, every party also holds a share of a corresponding MAC, computed with respect to a global secret key. These MACs are propagated and updated alongside the secret shares during the computation. When values are reconstructed or explicitly checked, the parties verify that the shares are consistent with their MACs; any deviation from the protocol will be detected with high probability, causing the protocol to abort. Maintaining and verifying these MACs introduces additional computational and communication overhead compared to the passive setting.

The protocols that we propose in Section \ref{sec:method} are sufficiently generic to be used in settings with passive or active adversaries. {\textcolor{black}{While in Section \ref{sec:results} we evaluate our protocols in settings where the number of computational parties $K=3$, they can be applied in settings with any number of computing parties $K \geq 2$}}. This is achieved by changing the underlying MPC scheme to align with the desired security setting, \textcolor{black}{without any modifications to the protocols in Section \ref{sec:method} themselves.} 
\textcolor{black}{One can for instance replace the 3-party schemes \cite{araki2016high,eerikson2020use}  that we use in Section \ref{sec:results} by an honest-majority 4-party computing scheme   \cite{dalskov2021fantastic}, a dishonest-majority 2-party scheme \cite{cryptoeprint:2018:482}, or an additive secret-sharing based MPC implementation that efficiently accommodates a larger number of parties (e.g.~$K=15$) \cite{Agarwal2019Protecting}.}

To give the reader a concrete understanding, we give a brief description of the \textit{semi-honest} MPC scheme based on replicated secret sharing with $K=3$ computing parties which we utilize \cite{araki2016high}. MPC works for inputs defined over a finite field or ring. When inputs for SDG algorithms are finite precision real numbers, as is the case in this paper, the data holders convert all of their inputs to fixed precision \cite{FC:CatSax10} and map these to integers modulo $q$, i.e., to the ring $\mathbb{Z}_q =\{0,1,\ldots,q-1\}$, with $q$ a power of 2. 
A private value $x$ in $\mathbb{Z}_q$ is secret shared among servers (parties) $S_1, S_2,$ and $S_3$ by selecting uniformly random shares $x_1, x_2, x_3 \in \mathbb{Z}_q$ such that 
$x_1 + x_2 +x_3 =  x \mod{q}$,
and giving $(x_1,x_2)$ to $S_1$, $(x_2,x_3)$ to $S_2$, and $(x_3,x_1)$ to $S_3$. We denote a secret sharing of $x$ by $[\![x]\!]$. 

From now on, servers will compute on the secret shares of $x$ rather than on $x$ itself. The three servers are capable of performing operations such as the addition of a constant, summing of secret-shared values, and multiplication by a publicly known constant by doing local computations on their respective shares. To multiply secret shared values $[\![x]\!]$ and $[\![y]\!]$, we can use the fact that $x \cdot y=(x_1 + x_2 +x_3)(y_1 + y_2 +y_3) \mod{q}$. This means that $S_1$ computes $z_1=x_1 \cdot y_1+x_1 \cdot y_2+x_2 \cdot y_1 \mod{q}$, $S_2$ computes $z_2=x_2 \cdot y_2+x_2 \cdot y_3+x_3 \cdot y_2 \mod{q}$, and $S_3$ computes $z_3=x_3 \cdot y_3+x_3 \cdot y_1+x_1 \cdot y_3 \mod{q}$. The next step is for the servers to obtain an additive secret sharing of $0$ by choosing random values $u_1,u_2,u_3$ such that $u_1 + u_2 +u_3 = 0 \mod{q}$. This can be done using pseudorandom functions. Each server $S_i$ then computes $v_i=z_i+u_i \mod{q}$. Finally, $S_1$ sends $v_1$ to $S_3$, $S_2$ sends $v_2$ to $S_1$, and $S_3$ sends $v_3$ to $S_2$. This allows the servers $S_1, S_2$, and $S_3$ to obtain the replicated secret shares $(v_1,v_2)$, $(v_2,v_3)$, and $(v_3,v_1)$, respectively, of the value $v=x \cdot y$.

In Section \ref{sec:results}, we evaluate our MPC solution for federated SDG in both passive and active threat models for an honest-majority setting with $K=3$ computational servers where one party can be corrupted by a semi-honest adversary \cite{araki2016high} or a malicious adversary \cite{eerikson2020use}. We assume point-to-point authenticated communication channels between parties. Detailed input/output semantics of the MPC primitives used in the paper are provided in Appendix A.

\subsection{Differential Privacy for Output Privacy} 
Differential privacy (DP) \cite{dwork2014algorithmic} is a privacy notion that offers theoretical bounds on how much a single individual can influence the outcome of some computation or algorithm. 
This means that even if an adversary has access to the output of the computation, they cannot determine whether any specific individual's data was included in the input dataset. DP bounds the difference in outputs resulting from two datasets $D$ and $D'$ differing in one record, set by privacy budget $\epsilon$. ($\epsilon$,$\delta$)-DP is a common relaxation of the stricter $\epsilon$-DP notion, permitting a small chance of mechanism failure quantified in $\delta$, i.e., with probability at most $\delta$, the mechanism may violate $\epsilon$-differential privacy. Formally, a randomized algorithm $M$ satisfies ($\epsilon$,$\delta$)-DP if and only if for any pair of neighboring datasets $D$ and $D'$, i.e., datasets that differ in only one record, and any $y \in Range(M)$:
\vskip -5pt
\begin{equation}
    Pr[M(D) = y] \leq e^\epsilon  Pr[M(D') = y]
    + \delta
\end{equation}

An ($\epsilon, \delta$)-DP algorithm $M$ is usually created out of an algorithm $M^*$ by adding noise 
that is proportional to the \textit{sensitivity} of $M^*$, 
in which the sensitivity reflects the maximum impact changing a single record in $D$ can have on the output of $M^*$. 

\subsection{Private-PGM for Synthetic Data Generation}

In this paper, we adapt the marginals-based SDG algorithm Private-PGM \cite{mckenna2019graphical} to federated settings. 
Private-PGM fits an undirected graphical model to differentially private noisy measurements of low-dimensional marginals, from which synthetic records are drawn by sampling the learned model. 

The method operates on a dataset $D$ with discrete features $x = \{x_1, x_2, \ldots, x_d\}$. Each feature $x_i$ has a discrete domain $\Omega_i$. For a subset $q$ of features, the marginal $\mu_q(D)$ counts, for each combination of values $t \in \Omega_q$, how many times that combination appears in $D$, where $\Omega_q = \prod_{x_i \in q} \Omega_i$.   

Private-PGM begins by computing DP marginals $\mu_q(D) + \mathcal{N}(0, \sigma_q I)$, where $\mathcal{N}$ is Gaussian noise with scale $\sigma_q$ determined based on specified DP parameters $(\epsilon,\delta)$. Then it estimates the joint marginal distribution that best explains all the noisy measurements. Specifically, following the implementation of Chen et al.~\cite{chen2024towards} for RNA-seq data, the algorithm first computes 1-way marginals $\mu_g$ capturing the count distribution of each gene $g$ and the label $\mu_y$, as well as 2-way marginals $\mu_{g,y}$ capturing the joint distributions between each gene $g$ and the label $y$. DP noise is then injected into these statistics based on the DP privacy budget. It then estimates the joint marginal distribution that best explains all the noisy measurements, while in parallel inferring the parameters of the graphical model through belief propagation on a junction tree. Finally, it reconstructs a synthetic dataset from these noisy measurements using a graphical-model-based inference engine. We refer to McKenna et al.~\cite{mckenna2019graphical} for more details.


\vspace{-10pt}
\section{Related Work}
\noindent
\textbf{Synthetic RNA-seq Data Generation.} SDG algorithms have previously been developed and applied for generation of synthetic RNA-seq data in the centralized setting, i.e., when all real data used to train the generator resides with one data holder. Existing research encompasses both generation of synthetic bulk RNA-seq data \cite{ozturk2026towards} 
and synthetic single cell RNA-seq data \cite{luo2024scdiffusion,Song2024}.
Regarding bulk RNA-seq data, which is the focus of this paper, Chen et al.~\cite{chen2024towards} produced a comprehensive comparison of SDGs, demonstrating that the marginal-based Private-PGM algorithm \cite{mckenna2019graphical} can generate bulk RNA-seq counts with high downstream utility while offering formal differential privacy guarantees. 
Its effectiveness was confirmed in the recent Critical Assessment of Massive Data Analysis (CAMDA) Health Privacy Challenge.
\footnote{\url{https://bipress.boku.ac.at/camda2025/competitions/camda-2025/}} At the conclusion of this international benchmarking competition in 2025, Private-PGM achieved the highest utility synthetic RNA-seq data (generated based on real TCGA data) of all submissions \cite{ozturk2026towards}. While there may still be room for future improvement, these results suggest that Private‑PGM is a strong starting point for developing MPC protocols to enable SDG in federated settings where data is distributed across multiple holders.  

\vspace{0.5em}
\noindent
\textbf{Federated Synthetic Data Generation.}
Research on secure SDG from distributed data sources, or federated SDG, is nascent. Most existing work focuses on image data and some on tabular data \cite{little2023federated}. Initial federated synthesis approaches adhere closely to the original federated learning (FL)  paradigm in which each client performs local neural network training and exchanges only parameter values instead of raw data with a central aggregator \cite{xin2022federated}. These approaches pay for privacy with a loss in utility compared to the centralized setting, and they are designed and evaluated for clients who each have thousands of instances available for local model training. This is in stark contrast with the scenarios that could actually benefit the most from federated SDG, provided the proper technical solutions are in place, such as the rare cancer setting in which a data holder may have data for only a handful of individuals.

Very little research has been done on 
training synthetic data generators  across encrypted data silos. Pentyala et al.~proposed the \texttt{CaPS} framework (Collaborative and Private Synthetic Data Generation) with the first MPC protocols for marginal-based tabular data generation \cite{pentyala2024caps}. Their method has since been shown to yield synthetic data with a substantially higher fidelity than a more traditional FL approach \cite{maddock2024flaim}. Experiments in \cite{maddock2024flaim, pentyala2024caps} are however performed in a so-called horizontal partitioning set-up in which it is assumed that each data holder (hospital) computes marginals for their own data on-site and subsequently sends encrypted shares of those marginals to MPC servers to start further processing within MPC. For generation of synthetic RNA-seq data with Private-PGM, however, we need the data holders to send encrypted shares of the gene expression levels (not the marginals) to the MPC servers so that the servers can execute an MPC protocol for quantile binning across all data. Computation of marginals is subsequently performed on secret sharings of the binned data, without decrypting it. RNA-seq data is highly dimensional, resulting in a large number of marginals that needs to be computed in this way. 
Our initial attempt to use \texttt{CaPS} for generation of synthetic RNA-seq data \cite{pentyala2024end} did not scale because of an inefficient implementation of the computation of the marginals over gene expression levels. Fu et al.~\cite{towardsfu2025} 
recently identified the inefficiencies of \texttt{CaPS} and proposed to address them by optimizing secure sort-and-count operations. In this paper, we present an alternative solution that lends itself well to RNA-seq data, namely replacing iterative ``per sample per gene'' computations by one dot product over a vector of all samples per gene. This leads to an efficient and effective solution \textcolor{black}{for synthetic RNA-seq generation. To the best of our knowledge, this is the first work to implement an MPC protocol for Private-PGM.}



\vspace{-10pt}

\section{MPC Protocols for Synthetic RNA-seq Data Generation}
\label{sec:method}
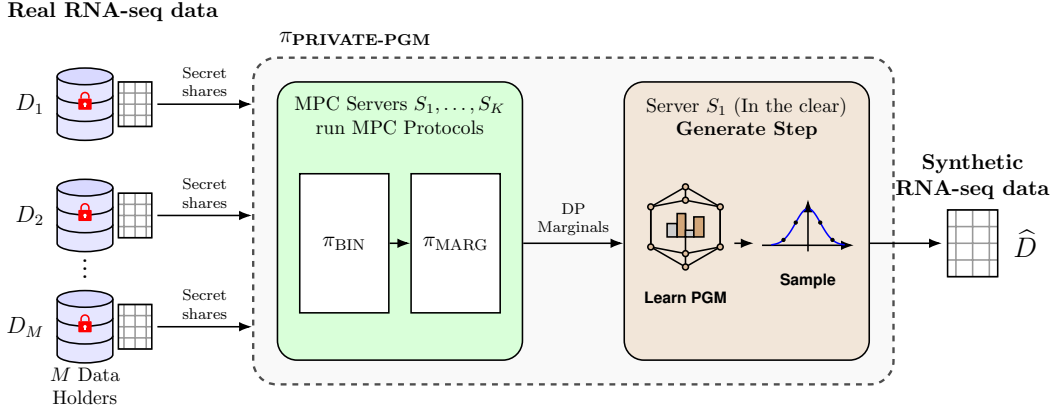
\begin{figure}
    \centering
    \resizebox{0.85\textwidth}{!}{ 
        \begin{tikzpicture}[
            >=Latex,
            font=\sffamily,
            innerbox/.style={
                draw, thick, fill=white, minimum width=1.6cm, minimum height=2.5cm,
                align=center, font=\normalsize
            },
            container/.style={
                draw, thick, rounded corners=10pt, minimum height=5.0cm, minimum width=4.4cm
            },
            rnagrid/.pic={
                \draw[step=0.2cm, gray!80, thick, fill=white] (-0.3,-0.4) grid (0.3, 0.4);
                \draw[thick] (-0.3,-0.4) rectangle (0.3, 0.4);
                \coordinate (-center) at (0,0);
                \coordinate (-west) at (-0.3,0);
                \coordinate (-east) at (0.3,0);
            },
            large rnagrid/.pic={
                \draw[step=0.3cm, gray!80, thick, fill=white] (-0.45,-0.6) grid (0.45, 0.6);
                \draw[thick] (-0.45,-0.6) rectangle (0.45, 0.6);
                \coordinate (-center) at (0,0);
                \coordinate (-west) at (-0.45,0);
                \coordinate (-east) at (0.45,0);
            },
            database/.pic={
                \draw[thick, fill=blue!10] (-0.5,0.5) -- (-0.5,-0.5) arc (180:360:0.5 and 0.15) -- (0.5,0.5);
                \draw[thick, fill=blue!15] (0,0.5) ellipse (0.5 and 0.15);
                \draw[thick] (-0.5,0.16) arc (180:360:0.5 and 0.15);
                \draw[thick] (-0.5,-0.16) arc (180:360:0.5 and 0.15);
                \begin{scope}[shift={(0,0)}]
                    \fill[red, rounded corners=0.5pt] (-0.12,-0.1) rectangle (0.12,0.1);
                    \draw[red, thick] (-0.07,0.1) -- (-0.07,0.14) arc (180:0:0.07) -- (0.07,0.1);
                    \fill[white] (0,0) circle (0.03);
                    \fill[white] (-0.015,-0.05) rectangle (0.015,0);
                \end{scope}
                \coordinate (-center) at (0,0);
                \coordinate (-east) at (0.5,0);
                \coordinate (-west) at (-0.5,0);
            }
        ]

        \pgfdeclarelayer{background}
        \pgfsetlayers{background,main}


        \node[container, fill=green!15] (mpc) at (0,0) {};
        \node[anchor=north, align=center, yshift=-0.15cm, font=\small] at (mpc.north) {MPC Servers $S_1, \ldots, S_K$\\run MPC Protocols};

        \node[container, fill=brown!20, right=1.8cm of mpc] (gen) {};
        \node[anchor=north, align=center, yshift=-0.15cm, font=\small] at (gen.north) {Server $S_1$\\\textbf{Generate Step}};

        \begin{pgfonlayer}{background}
            \node[draw, dashed, very thick, color=black!70, fill=gray!5, rounded corners=12pt, inner sep=12pt, fit=(mpc) (gen)] (pipgm) {};
            \node[anchor=south west, font=\large\bfseries, xshift=10pt, yshift=2pt] at (pipgm.north west) {$\pi_{\text{PRIVATE-PGM}}$};
        \end{pgfonlayer}


        \node[innerbox, anchor=center, xshift=-1.0cm, yshift=-0.4cm] (pibin) at (mpc.center) { $\pi_{\text{BIN}}$};
        \node[innerbox, anchor=center, xshift=1.0cm, yshift=-0.4cm] (pimarg) at (mpc.center) { $\pi_{\text{MARG}}$};
        \draw[->, thick] (pibin) -- (pimarg);

        \node[anchor=center, align=center] (pgm_step) at ([xshift=-1.1cm, yshift=-0.4cm]gen.center) {
            \begin{tikzpicture}[scale=0.6]
                \foreach \a in {30, 90, 150, 210, 270, 330} \coordinate (v\a) at (\a:1.2);
                \draw[thick] (v30) -- (v90) -- (v150) -- (v210) -- (v270) -- (v330) -- cycle;
                \draw[thick] (0, 0.8) -- (v90); \draw[thick] (0, 0.8) -- (v30); \draw[thick] (0, 0.8) -- (v150);
                \draw[thick] (0,-0.8) -- (v270); \draw[thick] (0,-0.8) -- (v210); \draw[thick] (0,-0.8) -- (v330);
                \draw[thick] (v150) -- (v210); \draw[thick] (v30) -- (v330);
                \draw[thick] (0,-0.8) -- (0,0.4);
                \foreach \a in {30, 90, 150, 210, 270, 330} \filldraw[fill=brown!50, draw=black, thick] (v\a) circle (2.5pt);
                \filldraw[fill=brown!50, draw=black, thick] (0, 0.8) circle (2.5pt);
                \filldraw[fill=brown!50, draw=black, thick] (0,-0.8) circle (2.5pt);
                \draw[fill=gray!40, thick] (-0.55,-0.3) rectangle (-0.25, 0.1);
                \draw[fill=brown!70, thick] (-0.25,-0.3) rectangle (0.0, 0.4);
                \draw[fill=gray!40, thick] (0.0,-0.3) rectangle (0.25, -0.1);
                \draw[fill=brown!70, thick] (0.25,-0.3) rectangle (0.55, 0.3);
            \end{tikzpicture}\\[3pt]
            \scriptsize \textbf{Learn PGM}
        };

        \node[anchor=center, align=center] (sample_step) at ([xshift=1.1cm, yshift=-0.4cm]gen.center) {
            \begin{tikzpicture}[scale=0.55]
                \draw[->, thick] (-1.5,0) -- (1.5,0);
                \draw[->, thick] (0,-0.2) -- (0,1.5);
                \draw[thick, blue, smooth, samples=50, domain=-1.2:1.2] plot(\x, {1.2*exp(-3*\x*\x)});
                \foreach \x in {-0.8, -0.4, 0, 0.4, 0.8} \filldraw[black] (\x, {1.2*exp(-3*\x*\x)}) circle (1.5pt);
            \end{tikzpicture}\\[3pt]
            \scriptsize \textbf{Sample}
        };

        \draw[->, thick] (pgm_step) -- (sample_step);


        \coordinate (db_mid_align) at ([xshift=-3.0cm]pipgm.west |- pibin.center);
        \coordinate (db_mid) at ([yshift=0.5cm]db_mid_align);

        \coordinate (db_top) at ([yshift=2.0cm]db_mid);
        \coordinate (db_bot) at ([yshift=-2.0cm]db_mid);

        \coordinate (title_y) at ([yshift=0.8cm]pipgm.north);
        
        \node[font=\bfseries, align=center] (real_title) at (db_mid |- title_y) [xshift=0.5cm] {Real RNA-seq data};

        \pic (db2) at (db_mid) {database};
        \pic (rna2) at ([xshift=0.9cm]db2-center) {rnagrid};
        \node[font=\large, left=0.1cm of db2-west] {$D_2$};

        \pic (db1) at (db_top) {database};
        \pic (rna1) at ([xshift=0.9cm]db1-center) {rnagrid};
        \node[font=\large, left=0.1cm of db1-west] {$D_1$};

        \pic (db3) at (db_bot) {database};
        \pic (rna3) at ([xshift=0.9cm]db3-center) {rnagrid};
        \node[font=\large, left=0.1cm of db3-west] {$D_M$};

        \path (db2-center) -- (db3-center) node[midway, font=\Large, yshift=0.15cm] {$\vdots$};
        \node[below=0.1cm of db3-center, align=center, font=\small, yshift=-0.5cm] {$M$ Data\\Holders};


        \coordinate (right_base) at ([xshift=1.4cm]pipgm.east);

        \pic (rna_synth) at (right_base |- sample_step.center) {large rnagrid};
        \node[font=\Large, right=0.15cm of rna_synth-east] {$\widehat{D}$};

        \node[font=\bfseries, align=center] (synth_title) at ([yshift=1.2cm]rna_synth-center) {Synthetic\\RNA-seq data};


        \foreach \i in {1,2,3} {
            \draw[->, thick] ([xshift=0.1cm]rna\i-east) -- node[above, font=\scriptsize, align=center] {Secret\\shares} (rna\i-east -| pipgm.west);
        }

        \draw[->, thick] (mpc.east |- pimarg.center) -- node[above, font=\scriptsize, align=center] {DP\\Marginals} (gen.west |- pimarg.center);

        \draw[->, thick] (gen.east |- rna_synth-west) -- ([xshift=-0.1cm]rna_synth-west);

        \end{tikzpicture}
    }
    
    \caption{Overview of the proposed $\piPGM$ solution for federated synthetic RNA-seq data generation. $M$ data holders each have a dataset ($D_1, D_2, \dots, D_M$) with RNA-seq data from patients. Each data holder encrypts its RNA-seq data and sends secret shares to a set of MPC servers $S_1, \dots , S_K$. The MPC servers run MPC protocols $\pi_{\text{BIN}}$ and $\pi_{\text{MARG}}$ to obtain encrypted marginal distributions estimated from the data \textit{while it stays encrypted}. 
    They subsequently run protocol $\piGAUSS$ to perturb the encrypted estimated marginals with noise to 
    provide differential privacy guarantees. Next the noisy marginals are decrypted by server $S_1$ and used to learn a probabilistic graphical model (PGM) that is subsequently sampled to generate a synthetic RNA-seq dataset $\widehat{D}$. {\textcolor{black}{The decryption step by $S_1$ can be done without breaking the method's DP guarantees because the marginals were already privatized in $\piGAUSS$.}}}
    \label{fig:overall_method}
\end{figure}
\paragraph{Formal Problem description.} We consider a federated scenario where each of $M$ data holders holds a local private dataset $D_i$. Each sample (i.e., each row) in $D_i$ consists of $d$ gene expression values as well as a label that denotes e.g.~the cancer type.
The data holders' goal is to leverage $\{D_1, D_2, \dots, D_M\}$ to generate synthetic data $\widehat{D}$.
This process must preserve both \textit{input privacy}, i.e.~no data holder reveals their raw records in $D_i$ to anyone, not even to a central server, and \textit{output privacy}, i.e., the generated $\widehat{D}$ does not leak sensitive information about the original samples in $\{D_1, D_2, \dots, D_M\}$. 

\paragraph{Protocol overview.} 
As illustrated in Fig.~\ref{fig:overall_method},
our proposed solution \footnote{\url{https://github.com/Filienko/Pro-GENE-GEN-MPC}} begins with each of the $M$ data holders creating secret shares of their respective dataset $D_i$ ($i=1, \ldots, M$) and sending the secret shares to $K$ computing servers. 
The servers proceed with joint execution of MPC protocol $\pi_{\text{PRIVATE-PGM}}$, following the pseudocode in Protocol \ref{alg:main_protocol}. 
We build upon the MPC primitives 
available in the MP-SPDZ library \cite{keller2020mp}.
For more details, 
see Appendix \ref{app:mpc}.

In Phase 1, the MPC servers concatenate the secret shares $[\![D_i]\!]$ from the $M$ data holders and transpose the resulting combined matrix (Lines 1--2). This enables column-wise data access. This replaces thousands of sequential network communication rounds with batched, Single Instruction, Multiple Data (SIMD) array operations, preventing extensive instruction-unrolling. Then, the servers separate the label column from the features. The first $d$ rows representing the gene expressions are assigned to the feature matrix $[\![D^T]\!]$ (Line 3), while the final row is split into a separate label array $[\![Y]\!]$ (Line 4). Isolating the labels from the gene expression features allows to vectorize the future computation of 2-way marginals between the two.

In Phase 2, following the preprocessing strategy in \cite{chen2024towards}, on Line 5 the MPC servers execute the sub-protocol $\pi_{\text{BIN}}$ for quantile binning to discretize the gene expression data to make it compatible with Private-PGM. Next, the MPC servers execute the sub-protocol $\pi_{\text{MARG}}$ for computing 1-way and 2-way marginals over the encrypted discretized data (Line 6), and $\pi_{\text{BATCH-GAUSS}}$ for perturbing the computed marginals with noise to provide differential privacy (DP) guarantees (Line 7). We describe each of these sub-protocols in more detail below. The noisy marginals computed in Phase 2 satisfy DP and can therefore be revealed (analogous to decryption) without compromising individual privacy. This follows from the \textit{post-processing property} of DP which states that any computation applied to the output of a DP method preserves the same privacy guarantees and cannot increase privacy leakage \cite{dwork2014algorithmic}.

In line with this, in Phase 3, all MPC servers reveal their secret shares of the noisy marginals (Line 8) and bin means (Line 9) by sending them to a designated data release server, $S_1$. Server $S_1$ combines the shares to obtain plaintext privatized marginals and bin means. In Phase 4, these are used to locally train a probabilistic graphical model to sample and generate the synthetic data $\widehat{D}$ (Lines 10-11). On Line 10, $S_1$ trains a PGM on the discrete data to  generate discrete synthetic data \cite{mckenna2019graphical}. On line Line 11, $S_1$ de-bins the discrete synthetic data using the released DP means to map it to its original continuous space.

\begin{myprotocol}
\caption{$\pi_{\text{PRIVATE-PGM}}$: Main MPC Protocol for Synthetic Data Generation}
\label{alg:main_protocol}
\footnotesize
\begin{algorithmic}[1]
\Require Secret shares of data $[\![D_i]\!]$, number $N_i$ of samples in $D_i$, number of data holders $M$, number of classes $C$, number of genes $d$, scale $\sigma$
\Ensure Synthetic dataset $\widehat{D}$

\Statex \textbf{$\triangleright$ Phase 1: Data Preparation}
\State $N \leftarrow \sum_{i=1}^{M} N_i$ \Comment{Total number of samples}
\State $[\![D_{\text{combined}}^T]\!] \leftarrow \text{Transpose}\left( \text{Concat}_{i=1}^M ([\![D_i]\!]) \right)$
\State $[\![D^T]\!] \leftarrow [\![D_{\text{combined}}^T[0 \dots d-1]]\!]$ \Comment{Gene features}
\State $[\![Y]\!] \leftarrow [\![D_{\text{combined}}^T[d]]\!]$ \Comment{Labels}

\Statex \textbf{$\triangleright$ Phase 2: Secure Quantile Binning and Computation of Noisy Marginals}
\State $[\![D^T]\!], [\![\mu_{\text{mean}}]\!] \leftarrow \pi_{\text{BIN}}([\![D^T]\!], N, d)$ \Comment{See Protocol \ref{alg:binning}}
\State $[\![\mu_g]\!], [\![\mu_y]\!], [\![\mu_{g,y}]\!] \leftarrow \pi_{\text{MARG}}([\![D^T]\!], [\![Y]\!], N, C, d)$ \Comment{See Protocol \ref{alg:marginals}}
\State $[\![\vec{\mu}_{\text{noisy}}]\!] \leftarrow \pi_{\text{BATCH-GAUSS}}([\![\mu_g]\!], [\![\mu_y]\!], [\![\mu_{g,y}]\!], \sigma)$ \Comment{See Protocol \ref{alg:batch_gauss}}

\Statex \textbf{$\triangleright$ Phase 3: Release of Differentially Private Marginals}
\State $\tilde{\mu}_g, \tilde{\mu}_y, \tilde{\mu}_{g,y} \leftarrow \pi_{\text{REVEAL}}([\![\vec{\mu}_{\text{noisy}}]\!])$ \Comment{Revealed to $S_1$}
\State $\tilde{\mu}_{\text{mean}} \leftarrow \pi_{\text{REVEAL}}([\![\mu_{\text{mean}}]\!])$   \Comment{Revealed to $S_1$}

\Statex \textbf{$\triangleright$ Phase 4: Synthetic Data Generation}
\State $\widehat{D}_{\text{discrete}} \leftarrow \text{Private-PGM}(\tilde{\mu}_g, \tilde{\mu}_y, \tilde{\mu}_{g,y})$ \Comment{Generate discrete data locally by $S_1$}
\State $\widehat{D} \leftarrow \text{Inverse-Binning}(\widehat{D}_{\text{discrete}}, \tilde{\mu}_{\text{mean}})$ \Comment{De-bin using DP means}
\State \Return $\widehat{D}$  
\end{algorithmic}
\end{myprotocol}

\paragraph{Secure quantile binning $\piBIN$.} 
The Private-PGM algorithm for SDG assumes that the features in the training data are categorical \cite{mckenna2019graphical}. To process RNA-seq data with Private-PGM, Chen et al.~suggested discretization of gene expression counts into 4 quantiles \cite{chen2024towards}. To perform such discretization efficiently over encrypted values in MPC, we adapt the binning logic from Pentyala et al.~\cite{pentyala2024end}. 
The protocol $\piBIN$ takes the gene expression matrix $[\![D^T]\!]$, maps each gene into 4 quantile bins,  and computes the mean value of each bin across all samples (Lines 2--15 in Protocol \ref{alg:binning}). For each gene $g$, the servers first extract the expression values into an array $[\![V_g]\!]$ (Line 3) and securely sort them (Line 4) to extract the quartile boundaries $[\![Q_0]\!], [\![Q_1]\!], [\![Q_2]\!]$ (Line 5). They then perform secure comparisons ($\pi_{\text{LT}}$) against these boundaries to create indicator arrays (Line 6). By combining these, a bin assignment array $[\![\vec{B}_g]\!]$ is computed, mapping every sample's value for gene $g$ to a discrete bin representing which quantile it falls into $\{0, 1, 2, 3\}$ (Line 7). For example, if a sample's value falls below the first quartile $[\![Q_0]\!]$, all three indicators yield $1$, mapping it to bin $3 - 1 - 1 - 1 = 0$. $[\![\vec{B}_g]\!]$ holds the secret shares of the newly binned features. Next, for each bin $b$ (Line 8), a Boolean mask $[\![\vec{m}_b]\!]$ is generated, representing whether the sample falls into bin $b$ (Line 9). Using these masks, the protocol securely counts the frequencies for each bin (Line 10). It then computes the mean expression of gene $g$ for samples in bin $b$ by multiplying the binary bin mask $[\![\vec{m}_b]\!]$ by the continuous expression values $[\![V_g]\!]$ to securely sum only the values of the samples belonging to bin $b$ (Line 11). This filtered sum is then divided by the total number of samples in that bin, $[\![\text{count}_b]\!]$, to yield the exact bin mean (Line 12). Before returning, the original gene expression values are replaced with their discrete bin indices (Line 14).

We optimize this process for MPC to minimize network latency and multiplicative depths, which otherwise slow down compilation and execution times. Unlike previous sequential approaches that process one sample at a time \cite{pentyala2024end}, we replace the iterative loops over each sample previously used to compute bin frequencies and means with vector dot products (Protocol \ref{alg:binning}, Lines 8--13). This collapses the original per-sample, per-gene iterative computations to just one dot product over a vector of $N$ samples per gene. 


\begin{myprotocol}
\caption{$\pi_{\text{BIN}}$: MPC Protocol for Quantile Binning}
\label{alg:binning}
\footnotesize
\begin{algorithmic}[1]
\Require Transposed secret-shared dataset $[\![D^T]\!]$, number of samples $N$, number of genes $d$
\Ensure Secret-shared binned dataset $[\![D^T]\!]$, secret-shared bin means $[\![\mu_{\text{mean}}]\!]$

\State $[\![\vec{1}_N]\!] \leftarrow$ Array of 1s of size $N$ 

\For{every gene $g \in \{0, \dots, d-1\}$}
    \State $[\![V_g]\!] \leftarrow [\![D^T[g]]\!]$ 
    
    \State $[\![S_g]\!] \leftarrow \pi_{\text{SORT}}([\![V_g]\!])$ 
    \State Extract quartiles: $[\![Q_0]\!], [\![Q_1]\!], [\![Q_2]\!] \leftarrow [\![S_g[\lfloor 0.25N \rfloor]]\!], [\![S_g[\lfloor 0.50N \rfloor]]\!], [\![S_g[\lfloor 0.75N \rfloor]]\!]$
    
    \State $[\![\vec{c}_0]\!] \leftarrow \pi_{\text{LT}}([\![V_g]\!], [\![Q_0]\!])$ ; $[\![\vec{c}_1]\!] \leftarrow \pi_{\text{LT}}([\![V_g]\!], [\![Q_1]\!])$ ; $[\![\vec{c}_2]\!] \leftarrow \pi_{\text{LT}}([\![V_g]\!], [\![Q_2]\!])$
    \State $[\![\vec{B}_g]\!] \leftarrow \vec{3} - [\![\vec{c}_2]\!] - [\![\vec{c}_1]\!] - [\![\vec{c}_0]\!]$   
    
    \For{$b \in \{0, 1, 2, 3\}$} 
        \State $[\![\vec{m}_b]\!] \leftarrow \pi_{\text{EQ}}([\![\vec{B}_g]\!], b)$
        \State $[\![\text{count}_b]\!] \leftarrow [\![\vec{m}_b]\!] \cdot \vec{1}_N$
        \State $[\![T]\!] \leftarrow \pi_{\text{DOT}}([\![\vec{m}_b]\!], [\![V_g]\!])$
        \State $[\![\mu_{\text{mean}}[g][b]]\!] \leftarrow \pi_{\text{DIV}}([\![T]\!], [\![\text{count}_b]\!])$
    \EndFor
    \State $[\![D^T[g]]\!] \leftarrow [\![\vec{B}_g]\!]$ 
\EndFor
\State \Return $[\![D^T]\!], [\![\mu_{\text{mean}}]\!]$
\end{algorithmic}
\end{myprotocol}

\paragraph{Secure computation of marginals $\pi_{\text{MARG}}$.}
The servers execute MPC protocol $\piMARG$ to compute the 1-way marginals (gene frequencies) and 2-way marginals (gene-label correlations). To achieve this efficiently on secret-shared data, we use indicator polynomials to convert a discrete bin index from $\{0, 1, 2, 3\}$ computed by $\pi_{\text{BIN}}$ into a one-hot encoded vector. This is necessary because, in MPC, servers cannot directly use secret-shared bin indices to access arrays without leaking the underlying data. By evaluating the indicator polynomials described below, the servers can construct the indicator vectors entirely through additions and multiplications, avoiding computationally expensive equality check operations, which significantly improves the efficiency of $\pi_{\text{MARG}}$. 
For a target bin $f \in \{0, 1, 2, 3\}$, the polynomial $G_f(x)$ is defined as:

\begin{center}
{{
$G_0(x) = \frac{(1-x)(2-x)(3-x)}{6};\ \  
G_1(x) = \frac{x(2-x)(3-x)}{2};\ \
G_2(x) = \frac{x(x-1)(3-x)}{2};\ \ 
G_3(x) = \frac{x(x-1)(x-2)}{6}$
}}
\end{center}

\noindent
By evaluating these polynomials, $G_f(x) = 1$ if $x = f$, and $0$ otherwise. 

This allows us to express the 2-way marginal count $\mu_{g,y}(f, c)$ for a gene $g$ at bin $f$ and class label $c$ as a sum of products. If $L_c(y_i)$ represents a Boolean indicator for whether the $i$-th sample belongs to class $c$, the 2-way marginal is computed as $\mu_{g,y}(f, c) = \sum_{i=1}^{N} G_f(x_i) \cdot L_c(y_i)$.

\begin{myprotocol}
\caption{$\pi_{\text{MARG}}$: MPC Protocol for Computing Marginals}
\label{alg:marginals}
\footnotesize
\begin{algorithmic}[1]
\Require Secret-shared binned dataset $[\![D^T]\!]$, target labels $[\![Y]\!]$, number of samples $N$, number of classes $C$, number of genes $d$
\Ensure Exact 1-way marginals $[\![\mu_g]\!], [\![\mu_y]\!]$, and 2-way marginals $[\![\mu_{g,y}]\!]$

\State $[\![\vec{1}_N]\!] \leftarrow$ Array of 1s of size $N$ 

\For{$c \in \{0, \dots, C-1\}$}
    \State $[\![\vec{L}_c]\!] \leftarrow \pi_{\text{EQ}}([\![Y]\!], c)$ 
    \State $[\![\mu_y[c]]\!] \leftarrow [\![\vec{L}_c]\!] \cdot \vec{1}_N$
\EndFor

\For{every gene $g \in \{0, \dots, d-1\}$}
    \State $[\![\vec{x}]\!] \leftarrow [\![D^T[g]]\!]$ 
    
    \State $[\![\vec{s}_1]\!] \leftarrow \vec{1} - [\![\vec{x}]\!]$ ; $[\![\vec{s}_{11}]\!] \leftarrow [\![\vec{x}]\!] - \vec{1}$ ; $[\![\vec{s}_2]\!] \leftarrow \vec{2} - [\![\vec{x}]\!]$ ; $[\![\vec{s}_{21}]\!] \leftarrow [\![\vec{x}]\!] - \vec{2}$ ; $[\![\vec{s}_3]\!] \leftarrow \vec{3} - [\![\vec{x}]\!]$
    
    \State $[\![\vec{G}_0]\!] \leftarrow \pi_{\text{MUL}}([\![\vec{s}_1]\!], [\![\vec{s}_2]\!], [\![\vec{s}_3]\!]) / 6$; 
    \State $[\![\vec{G}_1]\!] \leftarrow \pi_{\text{MUL}}([\![\vec{x}]\!], [\![\vec{s}_2]\!], [\![\vec{s}_3]\!]) / 2$; 
    \State $[\![\vec{G}_2]\!] \leftarrow \pi_{\text{MUL}}([\![\vec{x}]\!], [\![\vec{s}_{11}]\!], [\![\vec{s}_3]\!]) / 2$; 
    \State $[\![\vec{G}_3]\!] \leftarrow \pi_{\text{MUL}}([\![\vec{x}]\!], [\![\vec{s}_{11}]\!], [\![\vec{s}_{21}]\!]) / 6$
    
    \For{$f \in \{0, 1, 2, 3\}$}
    \State $[\![\mu_g[g][f]]\!] \leftarrow [\![\vec{G}_f]\!] \cdot \vec{1}_N$ \Comment{Local dot product}
\For{$c \in \{0, \dots, C-1\}$}
            \State $[\![\mu_{g,y}[g][f \cdot C + c]]\!] \leftarrow \pi_{\text{DOT}}([\![\vec{G}_f]\!], [\![\vec{L}_c]\!])$
        \EndFor
    \EndFor
\EndFor

\State \Return $[\![\mu_g]\!], [\![\mu_y]\!], [\![\mu_{g,y}]\!]$
\end{algorithmic}
\end{myprotocol}

First, the servers iterate over the $C$ target classes to generate Boolean indicator vectors $[\![\vec{L}_c]\!]$ for each class (Line 3 in Protocol \ref{alg:marginals}), summing them via dot product to calculate the exact 1-way marginals for the label array $[\![\mu_y]\!]$ (Line 4). Next, for each gene $g$ (Line 6), the servers extract the corresponding binned array $[\![\vec{x}]\!]$ (Line 7), calculate the intermediate values (Line 8), and compute the polynomials to generate the one-hot encoded arrays $[\![\vec{G}_0]\!]$ to $[\![\vec{G}_3]\!]$ for the four bins in $\{0,1,2,3\}$ (Line 9). Finally, for each target bin $f \in \{0,1,2,3\}$, the servers compute 1-way marginals of the gene by taking the dot product between the encoded array $[\![\vec{G}_f]\!]$ and the vector of ones (Line 11). To compute the 2-way marginals between the gene and the target labels, the servers evaluate our previously defined formula by computing dot products between $[\![\vec{G}_f]\!]$ and the label indicator vectors $[\![\vec{L}_c]\!]$ for each class $c$ (Lines 12--13).

This approach improves the compilation and execution speed of the protocol compared to an earlier attempt \cite{pentyala2024end}, which sequentially iterated over individual samples to compute the marginals and inject differential privacy (DP) noise element-wise. Such sequential processing resulted in a high number of repeated operations and network communication rounds, leading to slower execution speeds and large memory overhead. To resolve this, we apply a dot-product optimization: instead of sequentially iterating over individual samples, we compute the 1-way and 2-way marginal counts securely with vectorized dot products between the feature and the label indicator vectors (Line 11 for 1-way marginals and Line 13 for 2-way marginals).

\paragraph{Secure perturbation with noise for differential privacy with $\piGAUSS$.} We extend this vectorization to the DP noise phase as well. After the computing servers generate secret-shared Irwin-Hall approximated Gaussian noise (see  Protocol \ref{alg:irwin_hall} based on \cite{pentyala2024caps}), we flatten the secret-shared marginal vectors and inject this noise in a single vectorized addition in Protocol \ref{alg:batch_gauss}. Following Chen et al. \cite{chen2024towards}, we use a Rényi Differential Privacy (RDP) moments accountant \cite{mironov2017renyi} to compute the noise scale $\sigma$ based on the privacy budget $\epsilon$ (values specified in Section \ref{sec:results}) and $\delta = 10^{-5}$. On Line 4, the Irwin-Hall noise vectors are multiplied by $\sigma$ and added to the 1-way and 2-way marginal counts to guarantee DP.  This replaces thousands of element-wise network requests with one collective batched operation, resulting in significantly faster compilation.



\begin{myprotocol}
\caption{$\pi_{\text{BATCH-GAUSS}}$: MPC Batched Gaussian Addition}
\label{alg:batch_gauss}
\footnotesize
\begin{algorithmic}[1]
\Require Exact marginals $[\![\mu_g]\!], [\![\mu_y]\!], [\![\mu_{g,y}]\!]$, noise scale $\sigma$
\Ensure Flat array of noisy marginals $[\![\vec{\mu}_{\text{noisy}}]\!]$

\State $[\![\vec{\mu}_{\text{flat}}]\!] \leftarrow \text{Flatten}([\![\mu_g]\!], [\![\mu_y]\!], [\![\mu_{g,y}]\!])$
\State $M \leftarrow \text{length}([\![\vec{\mu}_{\text{flat}}]\!])$
\State $[\![\vec{\mathcal{N}}]\!] \leftarrow \pi_{\text{GAUSS}}(M)$ \Comment{See Appendix \ref{app:irwin_hall}} 
\State $[\![\vec{\mu}_{\text{noisy}}]\!] \leftarrow [\![\vec{\mu}_{\text{flat}}]\!] + \sigma \cdot [\![\vec{\mathcal{N}}]\!]$

\State \Return $[\![\vec{\mu}_{\text{noisy}}]\!]$
\end{algorithmic}
\end{myprotocol}

\paragraph{Privacy guarantees for the overall protocol.}
The \textit{input privacy} of Protocol 1 for federated SDG follows from our use of secure MPC primitives (see Appendix A) and the universal composition theorem, which states that an MPC protocol remains cryptographically secure when composed with other or the same MPC protocols \cite{canetti2000security}. Protocol 1 furthermore provides the same output privacy guarantees in the federated setting as those provided by Chen et al.~\cite{chen2024towards} in the centralized setting. As in Chen et al.~\cite{chen2024towards}, bin means are revealed in plaintext (Protocol 1, Line 9) and do not carry a formal DP guarantee. This reflects a limitation inherited from the original method rather than one introduced by our federated construction. While the disclosure of bin means may enable certain forms of information leakage and therefore merits further investigation, addressing or mitigating such leakage is orthogonal to the primary contribution of this work, which is demonstrating that the approach of Chen et al.~can be faithfully and securely realized in a federated setting using MPC.

\vspace{-10pt}
\section{Empirical Evaluation Results}
\label{sec:results}

\paragraph{Datasets.} We evaluate our approach  on four datasets across different cancer types. The first dataset, which we designate as ALL,  contains 1,181 samples across four leukemia subtypes 
\cite{warnat2020scalable}; it was used in a previous study of centralized SDG methods for genomic data \cite{chen2024towards}. The second, which we designate as AML, is a more recent dataset of Acute Myeloid Leukemia (AML) subtypes 
containing 1,224 bulk RNA-seq samples across 14 subtype labels \cite{severens2024mapping}. We also evaluate our method on two public benchmark datasets from the CAMDA 2025 competition on privacy-preserving genomic data generation \cite{ozturk2026towards}: TCGA-BRCA, a breast cancer dataset with 1,089 samples across five subtypes, and TCGA-COMBINED, a larger collection of 4,323 samples spanning ten cancer tissue types. 
Together, these datasets span diverse domains, class distributions, and label cardinalities, enabling a thorough evaluation of our method under conditions analogous to real-world federated rare disease settings.

We follow the preprocessing pipeline of Chen et al. (2024) \cite{chen2024towards}. We apply DESeq2 normalization \cite{love2014moderated} to minimize the influence of varying sequencing depths and RNA compositions, followed by gene filtering and ensuring log-normalization to account for the well-known skewness of RNA-seq counts. {\textcolor{black}{We implement uniform preprocessing of the datasets to establish a consistent evaluation environment and ensure a fair comparison between different datasets coming from various sources.
Developing a decentralized preprocessing and normalization algorithm for genomic data is an important next step that falls outside the primary scope of this paper. }}
The categorical labels are encoded as integers. After filtering, the datasets ALL, AML, TCGA-BRCA, and TCGA-COMB were reduced to 958, 1,000, 978, and 978 features, respectively. 


\paragraph{Experimental Setup.} To evaluate the fidelity and utility of the synthetic data, we divide each dataset into a training set $D_{train}$ and a held-out test set $D_{test}$ using a 80/20 data split. For our experiments across different numbers of features, we randomly subsample the genes to some number $X$. Next we distribute the samples in $D_{train}$ among $M$ data holders. Because all operations in $\piPGM$ are performed by the MPC servers over an encrypted version of the entire dataset $D_{train}$, the results in this section are agnostic to the value of $M$. 
We implemented $\piPGM$ in MP-SPDZ using both passive 3PC \cite{araki2016high} and active 3PC schemes \cite{eerikson2020use} for $K=3$ MPC servers, where the former represents a semi-honest adversary and the latter represents a malicious adversary.

\paragraph{Evaluation Metrics.} Following prior work \cite{chen2024towards, menziessynthetic__camda}, we assess synthetic data quality across multiple dimensions. \textit{Utility} is evaluated via a Train-on-Synthetic, Test-on-Real (TSTR) framework: we use the generated synthetic data to train a logistic regression (LR) model for classification of samples according to subtypes, and measure the accuracy of the trained LR model on the held-out test set $D_{test}$. 
\textit{Fidelity} is quantified by the Wasserstein distance between the continuous gene expression values 
in synthetic and real data. \textit{Biological value} is assessed by differential expression preservation (TPR of real differential expressed genes recovered in synthetic data). \textit{Output privacy}, following CAMDA \cite{ozturk2026towards}, is assessed by Distance to Closest Record (DCR), which is the mean Euclidean distance from each synthetic sample to its nearest real neighbor. Smaller DCR reflects more realistic looking data and hence potential privacy leakage.

\begin{figure}[t]
    \centering

    \newcommand{\epsfig}[3]{%
        \begin{subfigure}[b]{0.4\textwidth}
            \centering
            \begin{minipage}[c]{0.10\textwidth}
                \centering
            \end{minipage}%
            \begin{minipage}[c]{0.93\textwidth}
                \centering
                \includegraphics[width=\textwidth]{#1}
                \par\vspace{2pt}
            \end{minipage}
            \caption{#2}
            \label{#3}
        \end{subfigure}%
    }

    \epsfig{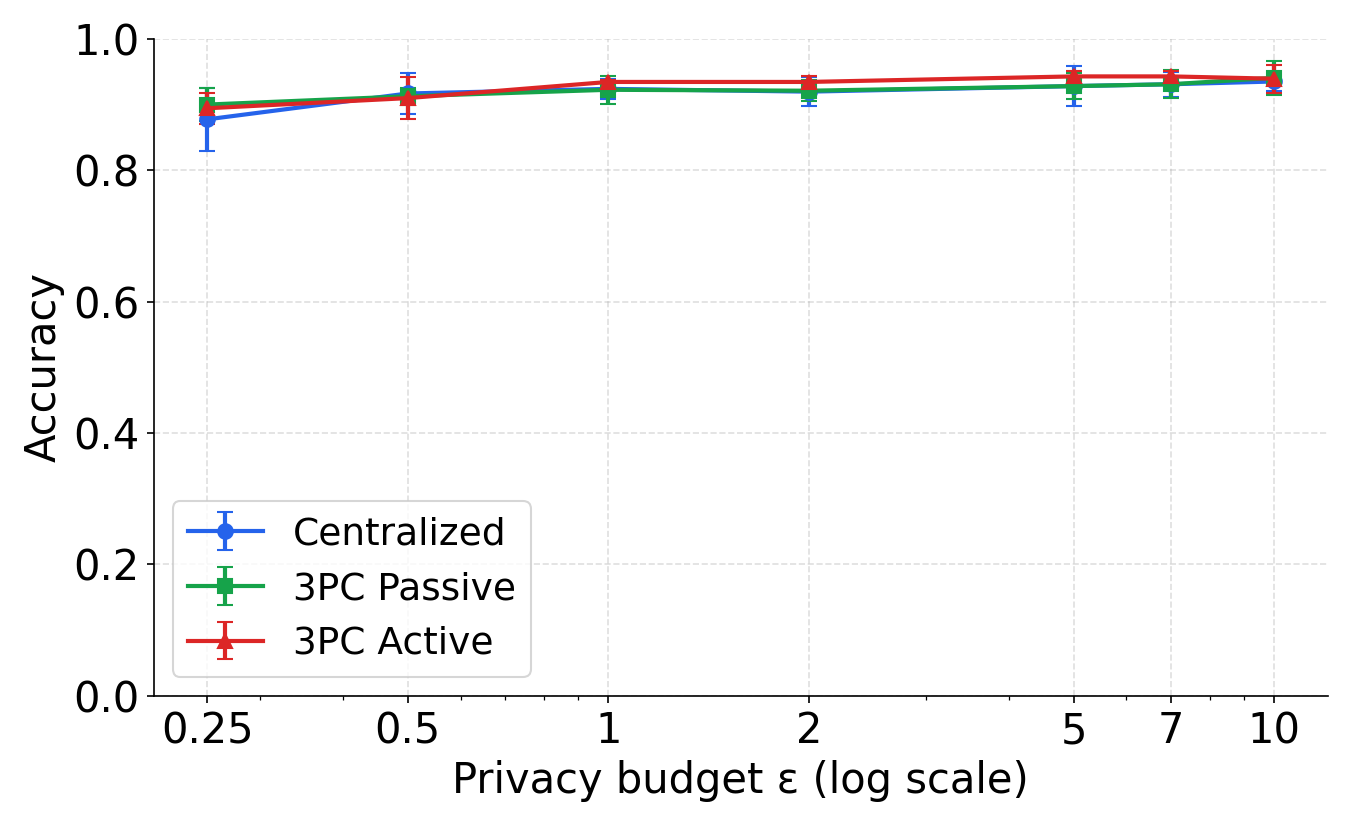}{ALL}{fig:eps_all}
    \hspace{0.02\textwidth}
    \epsfig{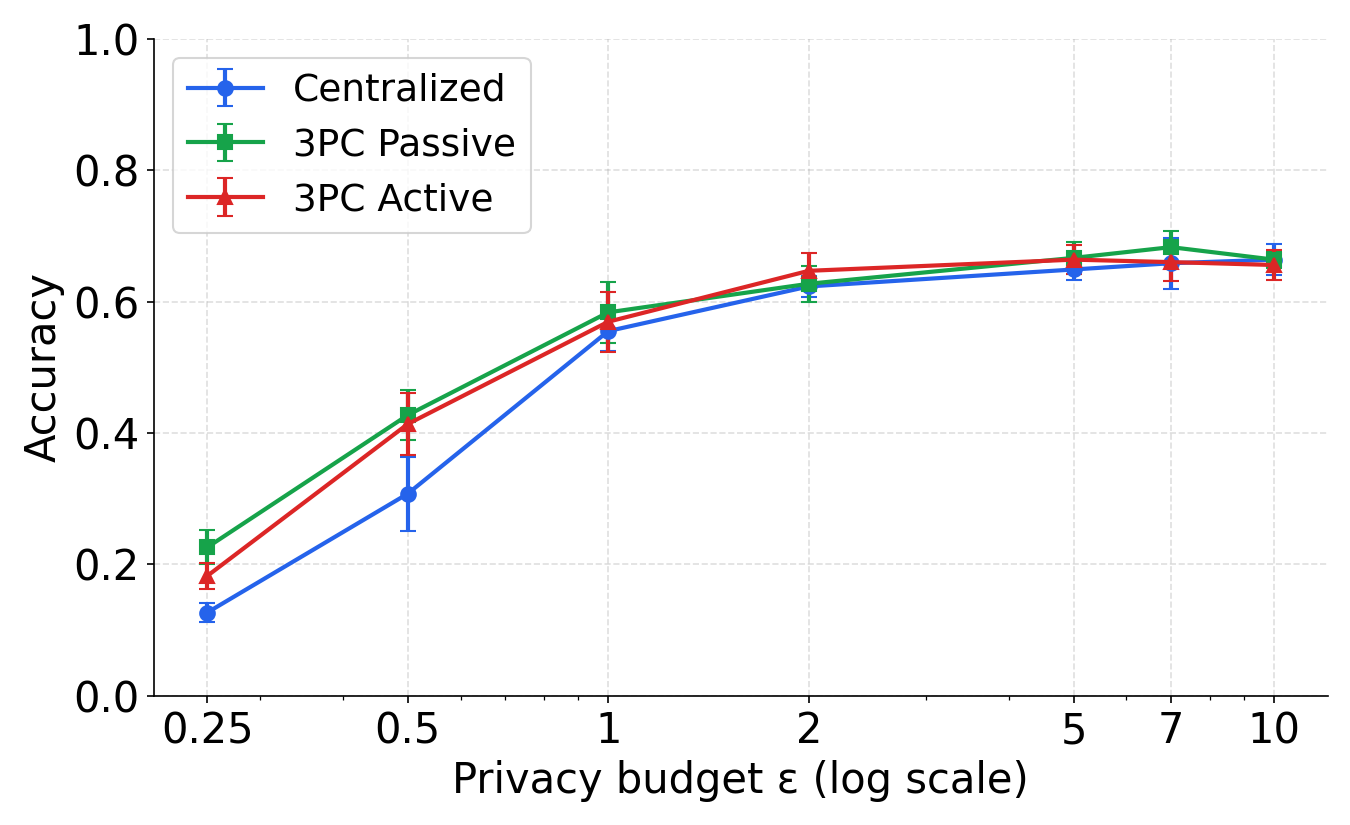}{AML}{fig:eps_aml}\\[6pt]
    \epsfig{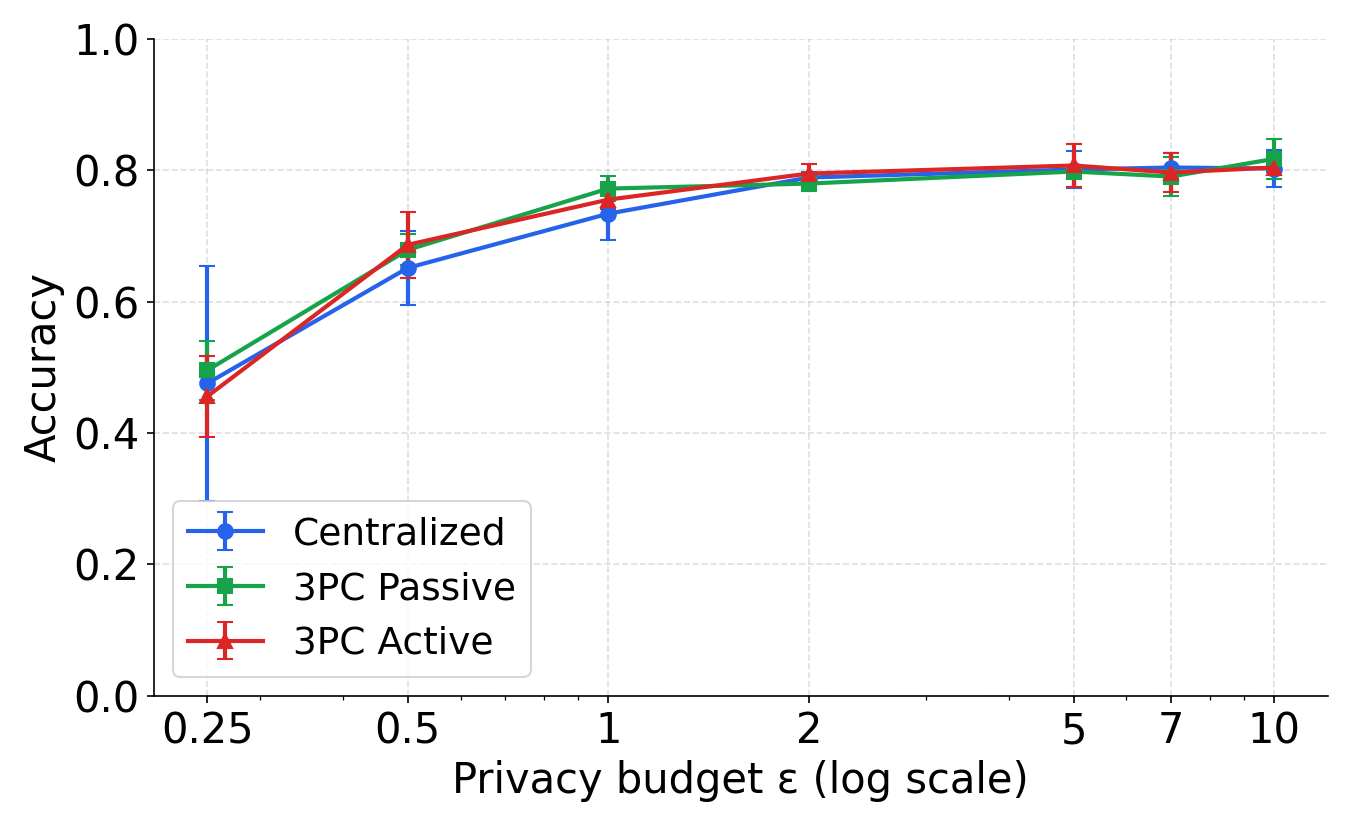}{TCGA-BRCA}{fig:eps_brca}
    \hspace{0.02\textwidth}
    \epsfig{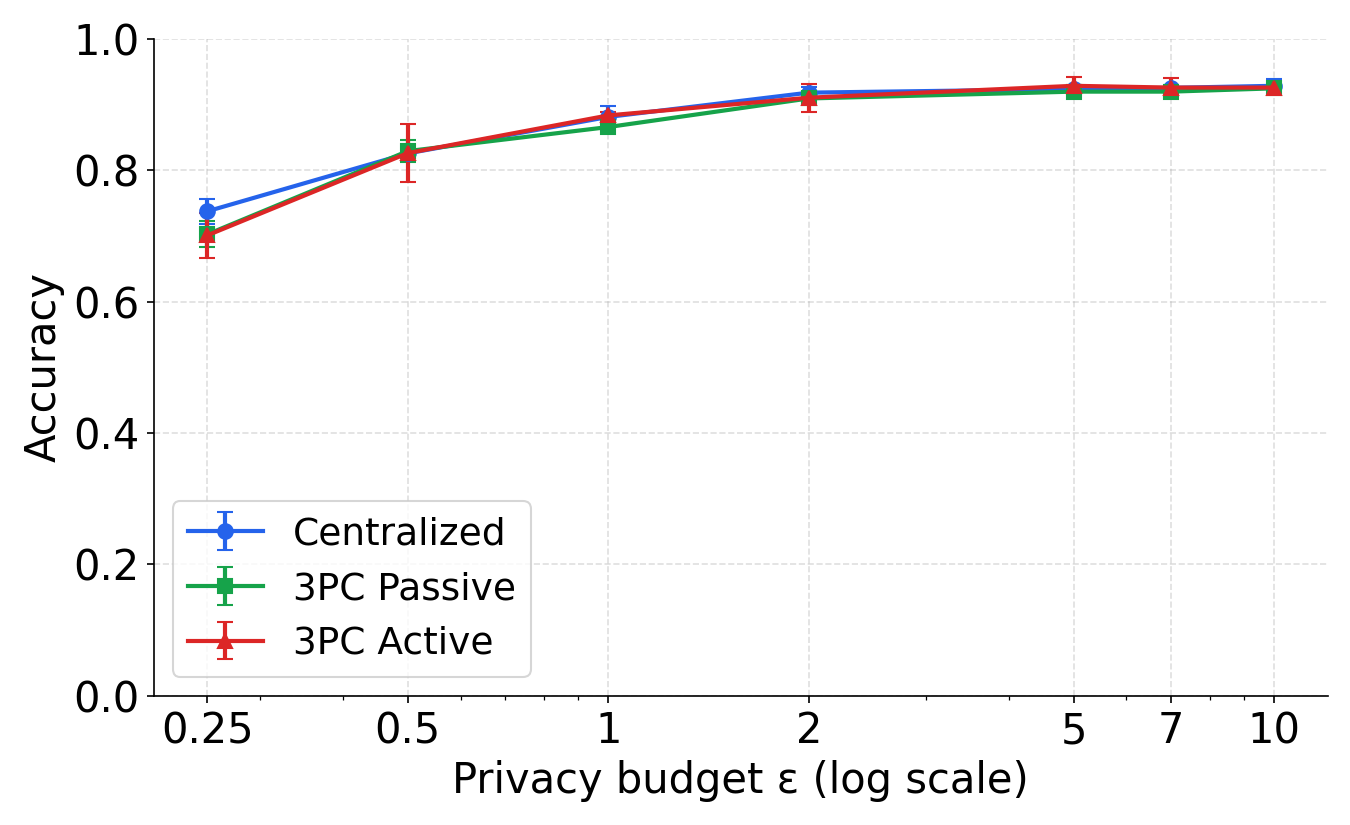}{TCGA-COMBINED}{fig:eps_comb}

    \caption{\textbf{Utility.} Classification accuracy of logistic regression models trained on synthetic data and evaluated on the real held-out test data across different privacy budgets $\epsilon$. Larger values of $\epsilon$ provide less output privacy. Our proposed \textbf{3PC passive} and \textbf{3PC active} federated SDG solutions achieve the same downstream utility as the centralized Private-PGM approach (only output privacy).}
    \label{fig:privacy_scaling_eps}
\end{figure}

\begin{table}[!h]
\centering
\small
\caption{ Empirical evaluation results on TCGA-COMBINED. Our proposed \textbf{3PC passive} and \textbf{3PC active} federated SDG solutions yield synthetic data with similar quality to the synthetic data generated in the centralized setting (output privacy only). All methods were evaluated at $\epsilon=10$.}
\label{tab:tcga_combined}
\begin{tabular}{clcccc}
\toprule
\textbf{Features} & \textbf{Protocol} & \textbf{Accuracy} & \textbf{DCR} & \textbf{DE-TPR} & \textbf{Wasserstein} \\
\midrule
\multirow{3}{*}{200}
 & Centralized  & 0.9318 & 12.322 & 0.9236 & 0.2708 \\
 & 3PC Active   & 0.9368 & 12.337 & 0.9278 & 0.2708 \\
 & 3PC Passive  & 0.9395 & 12.323 & 0.9250 & 0.2708 \\
\midrule
\multirow{3}{*}{600}
 & Centralized & 0.9561 & 24.872 & 0.8929 & 0.7194 \\
 & 3PC Active & 0.9514 & 26.842 & 0.8847 & 0.7191 \\
 & 3PC Passive & 0.9522 & 26.817 & 0.8836 & 0.7191 \\
\midrule
\multirow{3}{*}{800}
 & Centralized  & 0.9526 & 31.068 & 0.8688 & 0.6075 \\
 & 3PC Active   & 0.9520 & 30.967 & 0.8676 & 0.6072 \\
 & 3PC Passive  & 0.9558 & 30.992 & 0.8678 & 0.6079 \\
\midrule
\multirow{3}{*}{979}
 & Centralized  & 0.9576 & 34.622 & 0.8575 & 0.5489 \\
 & 3PC Active   & 0.9538 & 34.553 & 0.8584 & 0.5485 \\
 & 3PC Passive  & 0.9543 & 34.612 & 0.8564 & 0.5481 \\
\bottomrule
\end{tabular}
\end{table}

\paragraph{Results.} 
Our evaluation demonstrates that the proposed MPC solution achieves utility, fidelity, biological value,
and empirical privacy on par with the centralized baseline which provides only
output privacy. The results are consistent across all datasets and all recorded
metrics with some minor variation in performance which we attribute to DP noise. Tab.~\ref{tab:tcga_combined} reports all four metrics for
TCGA-COMBINED ($\epsilon=10$): downstream classification accuracy (utility), Wasserstein
distance on one-dimensional marginal distributions (fidelity), DE-TPR
(biological value), and DCR (output privacy). The results are averages across 3 runs with different seeds. 
Fig.~\ref{fig:privacy_scaling_eps} shows consistently high downstream classification accuracy
across privacy budgets $\varepsilon$ for all datasets across 100 features.
More results on these datasets across different privacy budgets are provided in
Appendix~\ref{app:more_results}.

\paragraph{Computational efficiency.}
Importantly, the input privacy-preserving benefits of our MPC solution are achieved at a low computational cost; as detailed in Tab.~\ref{tab:runtimes}, protocol compilation times are negligible (under 10 seconds), and execution times remain feasible across both  passive and active adversarial scenarios, even as the number of features increases. It is worth noting that the communication between parties for our protocols was simulated between the parties co-located on the same machine, and when deployed on a network with
separate machines connected via LAN/WAN would include additional communication overhead. \footnote{We ran our experiments on a machine with 128 GB RAM, 2 NVIDIA RTX A6000 GPUs, and an AMD Ryzen Threadripper PRO 5965WX (24 cores), using MP-SPDZ to simulate the multi-party computation environment. All parties were co-located on the same machine, communicating via a local communication channel.}

\begin{table}
\centering
\caption{End-to-end MPC protocol compile and execution time (seconds) by dataset and \#features, avg over 3 runs}
\label{tab:runtimes}
\footnotesize
\setlength{\tabcolsep}{5pt}
\begin{tabular}{ll *{4}{r} *{4}{r}}
\toprule
& & \multicolumn{4}{c}{\textbf{3PC Passive}} & \multicolumn{4}{c}{\textbf{3PC Active}} \\
\cmidrule(lr){3-6} \cmidrule(lr){7-10}
\textbf{Dataset} & \textbf{Time} & \textbf{200} & \textbf{500--600} & \textbf{800} & \textbf{959--1000}
                               & \textbf{200} & \textbf{500--600} & \textbf{800} & \textbf{959--1000} \\
\midrule
\multirow{2}{*}{BRCA \small($N$=1,089)}
  & Compile &  2.14 &  2.22 &  2.99 &   2.59 &   2.06 &   3.62 &   2.46 &   2.76 \\
  & Execute & 23.42 & 55.94 & 88.04 & 117.43 &  983.28 & 2735.95 & 4465.14 & 4800.20 \\
\addlinespace[4pt]
\multirow{2}{*}{ALL \small($N$=1,181)}
  & Compile &  2.13 &  2.41 &  2.63 &   2.83 &   2.22 &   3.36 &   3.14 &   2.89 \\
  & Execute & 25.20 & 57.76 & 90.92 & 117.30 & 1016.58 & 2915.58 & 4698.88 & 5627.41 \\

\addlinespace[4pt]
\multirow{2}{*}{AML \small($N$=1,224)}
  & Compile &  3.31 &  3.84 &  4.51 &   4.64 &   3.28 &   3.84 &   6.28 &   4.68 \\
  & Execute & 25.67 & 61.92 & 100.09 & 133.49 & 1077.39 & 2627.35 & 7013.80 & 5641.11 \\
\addlinespace[4pt]
\multirow{2}{*}{COMB \small($N$=4,323)}
  & Compile &  7.49 &  8.20 &  8.27 &   8.64 &   7.64 &   8.03 &   8.35 &   8.99 \\
  & Execute & 118.10 & 249.35 & 324.88 & 426.92 & 4337.63 & 12758.33 & 16330.31 & 20792.39 \\
\bottomrule
\end{tabular}
\begin{tablenotes}
\footnotesize
\item Feature counts 200/500/800/959--1000 for ALL, BRCA, AML; 200/600/800/979 for COMB. 3PC passive = semi-honest (ring) \cite{araki2016high}; 3PC active = malicious (mal-rep-ring) \cite{eerikson2020use}. $N$ is the number of samples in the dataset. 
\end{tablenotes}
\end{table}

With the linear scaling of protocol runtimes observed in Tab.~\ref{tab:runtimes}, we extrapolate that applying our method in the ``active" adversary threat model to full transcriptomic settings (20,000+ genes) would require roughly 6 days of computations on the same hardware for a dataset of the size of TCGA-COMBINED. For more details on the per-protocol runtimes, see Appendix~\ref{app:comm_cost}.
However, response time is less critical in this context, as synthetic data generation is inherently an offline process more akin to model training than to latency-sensitive inference.
We believe that even in the active setting, this runtime is still a tolerable one-time cost, given the benefits of privacy-preserving distributed synthetic data generation. Furthermore, it could be improved with a more powerful set of computational servers. It is also worth noting that our protocols are consistently faster and more computationally  efficient than the current state-of-the-art algorithm, as demonstrated in Appendix \ref{app:comparisons}.}

\vspace{-10pt}

\section{Conclusion}
We introduced an efficient MPC-based solution for privacy-preserving generation of synthetic bulk RNA-seq data in federated settings. By carefully designing MPC protocols, we enabled Private-PGM to run on distributed, encrypted, high-dimensional genomics data. Leveraging vectorized MPC primitives, our approach attains utility and biological fidelity comparable to centralized baselines while remaining computationally practical under both passive and active adversary models. Empirical results across multiple real-world cancer datasets demonstrate that high-quality synthetic genomic data can be generated jointly across institutions without disclosing raw patient records. We view this work as a concrete step toward alleviating persistent data-sharing barriers in genomics and enabling secure, scalable collaboration in rare disease and multi-site biomedical research.

\paragraph{Acknowledgments}
This research was supported by the UW Tacoma Founders Endowment, and the UW eScience Institute.
This material is based upon work supported by the National Science Foundation under Grant No.~2451163 and No.~2523406, and by NSF NAIRR 240485 (Cloudbank AWS, TACC Vista) and NSF NAIRR 240091 (TACC Frontera). This research was, in part, funded by the National Institutes of Health (NIH) Agreement No. 1OT2OD032581. The views and conclusions contained in this document are those of the authors and should not be interpreted as representing the official policies, either expressed or implied, of the NIH. The authors thank the reviewers for their valuable and detailed feedback.


\paragraph{Data and Software Availability}
The software developed for this study and data to run it on is publicly available at \url{https://github.com/Filienko/Pro-GENE-GEN-MPC}.

\bibliography{references}

@article{ozturk2026towards,
  title={{Towards Useful and Private Synthetic Omics: Community Benchmarking of Generative Models for Transcriptomics Data}},
  author={{\"O}zt{\"u}rk, Hakime and Afonja, Tejumade and J{\"a}lk{\"o}, Joonas and Binkyte, Ruta and Rodriguez-Mier, Pablo and Lobentanzer, Sebastian and Wicks, Andrew and Kreuer, Jules and Ouaari, Sofiane and Pfeifer, Nico and others},
  journal={bioRxiv},
  pages={2026--03},
  year={2026}
}

@article{warnat2020scalable,
  title={{Scalable Prediction of Acute Myeloid Leukemia
    Using High-Dimensional Machine Learning and Blood Transcriptomics}},
  author={Warnat-Herresthal, Stefanie and Perrakis, Konstantinos and Taschler, Bernd and Becker, Matthias and Ba{\ss}ler, Kevin and Beyer, Marc and G{\"u}nther, Patrick and Schulte-Schrepping, Jonas and Seep, Lea and Klee, Kathrin and others},
  journal={{iScience}},
  volume={23},
  number={1},
  year={2020},
  publisher={{Elsevier}}
}

@article{severens2024mapping,
	title        = {{Mapping AML heterogeneity-multi-cohort transcriptomic analysis identifies novel clusters and divergent ex-vivo drug responses}},
	author       = {Severens, Jeppe F and Karakaslar, E Onur and van der Reijden, Bert A and S{\'a}nchez-L{\'o}pez, Elena and van den Berg, Redmar R and Halkes, Constantijn JM and van Balen, Peter and Veelken, Hendrik and Reinders, Marcel JT and Griffioen, Marieke and others},
	year         = 2024,
	journal      = {{Leukemia}},
	publisher    = {Nature Publishing Group UK London},
	volume       = 38,
	number       = 4,
	pages        = {751--761}
}

@inproceedings{cryptoeprint:2018:482,
  title={{SPD}$\mathbb{Z}_{2^k}$: Efficient {MPC} mod $2^{k}$ for Dishonest Majority},
  author={Cramer, Ronald and Damg{\aa}rd, Ivan and Escudero, Daniel and Scholl, Peter and Xing, Chaoping},
  booktitle={Annual International Cryptology Conference},
  pages={769--798},
  year={2018},
  organization={Springer}
}

@article{Agarwal2019Protecting,
author = {Agarwal, A. and Dowsley, R. and McKinney, N. and Wu, D. and Lin, C.-T. and De Cock, M. and Nascimento, A.},
title = {Protecting Privacy of Users in Brain-Computer Interface Applications},
journal = {IEEE Transactions on Neural Systems and Rehabilitation Engineering},
volume = {27},
number = {8},
pages = {1546--1555},
year = {2019}
}

@inproceedings{menziessynthetic__camda,
	title = {Synthetic data generation for bulk {RNA}-seq data: a {CAMDA} health challenge analysis},
	booktitle    = {{Annual International Conference on Critical Assessment of Massive Data Analysis, collocated with ISMB/ECCB}},
	author = {Menzies, Shane and Pentyala, Sikha and Filienko, Daniil and Golob, Steven and Banerjee, Jineta and Foschini, Luca and De Cock, Martine},
	year = {2025}
}

@inproceedings{pentyala2024end,
  title={End to End Collaborative Synthetic Data Generation},
  author={Pentyala, Sikha and Sitaraman, Geetha and Claar, Trae and De Cock, Martine},
  booktitle={AAAI-25 Workshop on Privacy-Preserving Artificial Intelligence},
  journal={arXiv preprint arXiv:2412.03766},
  year={2024}
}

@article{luo2024scdiffusion,
  title={{scDiffusion:} conditional generation of high-quality single-cell data using diffusion model},
  author={Luo, Erpai and Hao, Minsheng and Wei, Lei and Zhang, Xuegong},
  journal={Bioinformatics},
  volume={40},
  number={9},
  pages={btae518},
  year={2024}
}

@inproceedings{pentyala2024caps,
	title        = {{{CaPS:} Collaborative and Private Synthetic Data Generation from Distributed Sources}},
	author       = {Sikha Pentyala and Mayana Pereira and Martine {De Cock}},
	year         = 2024,
	booktitle    = {{International Conference on Machine Learning}},
	series       = {PMLR},
	volume       = {235},
    pages        = {40397-40413}
}

@inproceedings{mironov2017renyi,
  title={{R{\'e}nyi Differential Privacy}},
  author={Mironov, Ilya},
  booktitle={30th Computer Security Foundations Symposium},
  pages={263--275},
  year={2017},
  organization={IEEE}
}

@inproceedings{towardsfu2025,
  title     = {{Towards Vertically Distributed Differentially Private Synthetic Data Generation}},
  author    = {Fu, Yucheng and Gu, Tiaoyao and Shi, Elaine and Wang, Tianhao},
  booktitle = {{Theory and Practice of Differential Privacy Workshop Series}},
  year      = {2025}
}

@book{CDN2015,
	title        = {{Secure Multiparty Computation and Secret Sharing}},
	author       = {Ronald Cramer and Ivan Damg{\aa}rd and Jesper Buus Nielsen},
	year         = 2015,
	publisher    = {Cambridge University Press}
}

@article{love2014moderated,
  title={{Moderated estimation of fold change and dispersion for RNA-seq data with DESeq2}},
  author={Love, Michael I and Huber, Wolfgang and Anders, Simon},
  journal={{Genome Biology}},
  volume={15},
  number={12},
  pages={550},
  year={2014},
  publisher={Springer}
}

@article{watson2023delivering,
  title={Delivering on {NIH} data sharing requirements: avoiding Open Data in Appearance Only},
   author={Watson, Hope and Gallifant, Jack and Lai, Yuan and Radunsky, Alexander P and Villanueva, Cleva and Martinez, Nicole and Gichoya, Judy and Huynh, Uyen Kim and Celi, Leo Anthony},
  journal={BMJ Health \& Care Informatics},
  volume={30},
  number={1},
  year={2023},
  publisher={BMJ Publishing Group}
}

@inproceedings{hu2023sok,
  title={{SoK}: Privacy-Preserving Data Synthesis},
  author={Hu, Yuzheng and Wu, Fan and Li, Qinbin and Long, Yunhui and Garrido, Gonzalo and Ge, Chang and Ding, Bolin and Forsyth, David and Li, Bo and Song, Dawn},
  booktitle={IEEE Symposium on Security and Privacy (SP)},
  pages={4696-4713},
  year={2024}
}

@article{golob2024privacy,
  title = {Privacy Vulnerabilities in Marginals-based Synthetic Data},
	booktitle = {2025 {IEEE} {Conference} on {Secure} and {Trustworthy} {Machine} {Learning} ({SaTML})},
	author = {Golob, Steven and Pentyala, Sikha and Maratkhan, Anuar and De Cock, Martine},
	year = {2025},
	pages = {977--995}
}

@inproceedings{keller2020mp,
  title={{MP-SPDZ: A Versatile Framework for Multi-Party Computation}},
  author={Keller, Marcel},
  booktitle={{Proceedings of the 2020 ACM SIGSAC Conference on Computer and Communications Security}},
  pages={1575--1590},
  year={2020}
}

@inproceedings{chen2020gan,
  title={Gan-leaks: A taxonomy of membership inference attacks against generative models},
  author={Chen, Dingfan and Yu, Ning and Zhang, Yang and Fritz, Mario},
  booktitle={Proceedings of the 2020 ACM SIGSAC conference on computer and communications security},
  pages={343--362},
  year={2020}
}

@article{hilprecht2019monte,
  title={Monte carlo and reconstruction membership inference attacks against generative models},
  author={Hilprecht, Benjamin and H{\"a}rterich, Martin and Bernau, Daniel},
  journal={Proceedings on Privacy Enhancing Technologies},
  year={2019}
}

@article{hayes2019logan,
  author  = {Hayes, Jamie and Melis, Luca and Danezis, George and De Cristofaro, Emiliano},
  title   = {LOGAN: Membership Inference Attacks Against Generative Models},
  journal = {Proceedings on Privacy Enhancing Technologies},
  volume  = {2019},
  number  = {1},
  pages   = {133--152},
  year    = {2019},
  doi     = {10.2478/popets-2019-0008}
}

@inproceedings{eerikson2020use,
  title={{Use Your Brain! Arithmetic 3PC for Any Modulus with Active Security}},
  author={Eerikson, Hendrik and Keller, Marcel and Orlandi, Claudio and Pullonen, Pille and Puura, Joonas and Simkin, Mark},
  booktitle={{1st Conference on Information-Theoretic Cryptography}},
  year={2020},
  pages =	{5:1--5:24},
  series =	{Leibniz International Proceedings in Informatics (LIPIcs)},
  volume =	{163}
}

@article{fu2026revisiting,
  title={Revisiting Continuous Noise Sampling for Multi-Party Differential Privacy},
  author={Fu, Yucheng and Wang, Tianhao},
  journal={arXiv preprint arXiv:2608.27766},
  year={2026}
}

@inproceedings{araki2016high,
  title={{High-throughput semi-honest secure three-party computation with an honest majority}},
  author={Araki, Toshinori and Furukawa, Jun and Lindell, Yehuda and Nof, Ariel and Ohara, Kazuma},
  booktitle={{Proceedings of the 2016 ACM SIGSAC Conference on Computer and Communications Security}},
  pages={805--817},
  year={2016}
}

@article{dwork2014algorithmic,
  title   = {The algorithmic foundations of differential privacy},
  author  = {Dwork, Cynthia and Roth, Aaron},
  journal = {Foundations and Trends{\textregistered} in Theoretical Computer Science},
  volume  = {9},
  number  = {3--4},
  pages   = {211--407},
  year    = {2014}
}

@article{canetti2000security,
  title={{Security and Composition of Multiparty Cryptographic Protocols}},
  author={Canetti, Ran},
  journal={{Journal of Cryptology}},
  volume={13},
  number={1},
  pages={143--202},
  year={2000},
  publisher={Springer}
}

@inproceedings{chen2024towards,
 title={Towards Biologically Plausible and Private Gene Expression Data Generation},
  author={Chen, Dingfan and Oestreich, Marie and Afonja, Tejumade and Kerkouche, Raouf and Becker, Matthias and Fritz, Mario},
  booktitle={Proceedings in Privacy-Enhancing Technologies},
  volume={2},
  pages={531-554},
  year={2024}
}

@inproceedings{mckenna2019graphical,
  title = 	 {{Graphical-model based estimation and inference for differential privacy}},
  author =       {McKenna, Ryan and Sheldon, Daniel and Miklau, Gerome},
  booktitle = 	 {{Proceedings of the 36th International Conference on Machine Learning}},
  pages = 	 {4435--4444},
  year = 	 {2019},
  volume = 	 {97},
  series =    {PMLR}
}

@article{mckenna2022aim,
  title={{AIM:} An Adaptive and Iterative Mechanism for Differentially Private Synthetic Data},
  author={McKenna, Ryan and Mullins, Brett and Sheldon, Daniel and Miklau, Gerome},
  journal={Proceedings of the VLDB Endowment},
  volume={15},
  number={11},
  pages={2599--2612},
  year={2022}
}

@article{vanbreugel2023membershipinferenceattackssynthetic,
  title={Membership inference attacks against synthetic data through overfitting detection},
  author={van Breugel, Boris and Sun, Hao and Qian, Zhaozhi and van der Schaar, Mihaela},
  journal={AISTATS},
  year={2023},
  pages={3493--3514}
}

@article{oestreich2021privacy,
  title={Privacy considerations for sharing genomics data},
  author={Oestreich, Marie and Chen, Dingfan and Schultze, Joachim L and Fritz, Mario and Becker, Matthias},
  journal={EXCLI Journal},
  volume={20},
  year={2021},
  pages = {1243--1260}
}

@misc{NIH2025,
author={NIH},
title={Request for Information on Responsibly Developing and Sharing Generative Artificial Intelligence Tools Using {NIH} Controlled Access Data },
number={NOT-OD-25-118},
howpublished={Available at: grants.nih.gov/grants/guide/notice-files/NOT-OD-25-118.html},
year={2025}
}

@inproceedings{FC:CatSax10,
  author    = {Catrina, O. and
               Saxena, A.},
  title     = {Secure Computation with Fixed-Point Numbers},
  booktitle = {14th International Conference on Financial Cryptography and Data Security},
  series    = {Lecture Notes in Computer Science},
  volume    = {6052},
  pages     = {35--50},
  publisher = {Springer},
  year={2010},
}

@inproceedings{dalskov2021fantastic,
  title={Fantastic four: Honest-majority four-party secure computation with malicious security},
  author={Dalskov, Anders and Escudero, Daniel and Keller, Marcel},
  booktitle={{USENIX} 2021},
  pages={2183--2200},
  year={2021}
}

@Article{Song2024,
author={Song, Dongyuan
and Wang, Qingyang
and Yan, Guanao
and Liu, Tianyang
and Sun, Tianyi
and Li, Jingyi Jessica},
title={{scDesign3} generates realistic in silico data for multimodal single-cell and spatial omics},
journal={Nature Biotechnology},
year={2024},
day={01},
volume={42},
number={2},
pages={247-252}
}

@misc{NIH-NIST2024,
author={NIH},
title={Implementation Update for Data Management and Access Practices Under the Genomic Data Sharing Policy},
number={NOT-OD-24-157},
howpublished={Available at: grants.nih.gov/grants/guide/notice-files/NOT-OD-24-157.html},
year={2024}
}

@article{xin2022federated,
  title={Federated synthetic data generation with differential privacy},
  author={Xin, Bangzhou and Geng, Yangyang and Hu, Teng and Chen, Sheng and Yang, Wei and Wang, Shaowei and Huang, Liusheng},
  journal={Neurocomputing},
  volume={468},
  pages={1-10},
  year={2022}
}

@article{little2023federated,
  title={Federated learning for generating synthetic data: a scoping review},
  author={Little, Claire and Elliot, Mark and Allmendinger, Richard},
  journal={International Journal of Population Data Science},
  volume={8},
  number={1},
  year={2023},
  pages={2158}
}

@inproceedings{maddock2024flaim,
  title={{FLAIM: AIM}-based synthetic data generation in the federated setting},
  author={Maddock, Samuel and Cormode, Graham and Maple, Carsten},
  booktitle={Proceedings of the 30th ACM SIGKDD Conference on Knowledge Discovery and Data Mining},
  pages={2165--2176},
  year={2024}
}

\appendix
\onecolumn

\section{MPC Primitives}\label{app:mpc}
\noindent
In our proposed MPC protocols for federated synthetic data generation, we build upon the cryptographic primitives available in the MP-SPDZ \cite{keller2020mp} library. We use the following primitives as sub-protocols:

\begin{itemize}
\item Secure random number generation $\pi_{\text{RAND}}$: The parties securely generate a secret sharing $[\![r]\!]$ of a uniformly distributed random number $r$ via the bitwise XOR of locally generated random bits.

\item Secure sorting $\pi_{\text{SORT}}$: At the start of this protocol, the parties have a secret-shared array of $n$ elements $[\![\mathbf{x}]\!]$; at the end, they have a secret-shared array $[\![\mathbf{x}']\!]$ where the elements from $\mathbf{x}$ are sorted according to a specified order, without revealing the underlying values or the resulting data permutation to any party.


\item Secure multiplication $\pi_{\text{MUL}}$: At the start of this protocol, the parties have secret-shared values $[\![a]\!]$ and $[\![b]\!]$; at the end, they have a secret sharing of the product $[\![c]\!]$, where $c = a \cdot b$. 

\item Secure division $\pi_{\text{DIV}}$: At the start of this protocol, the parties have secret-shared values $[\![a]\!]$ and $[\![b]\!]$; at the end, they have a secret sharing of the quotient $[\![c]\!]$, where $c = a / b$.

\item Secure dot product $\pi_{\text{DOT}}$: At the start of this protocol, the parties have secret shares of two vectors $[\![\mathbf{a}]\!]$ and $[\![\mathbf{b}]\!]$ of length $n$; at the end, they have a secret sharing of their dot product $[\![c]\!]$, where $c = \sum_{i=1}^{n} a_i b_i$. The element-wise products are summed locally by the parties before a single secure truncation is applied to the final result, which reduces the overall truncation cost compared to standard element-wise multiplication.

\item Secure equality test $\pi_{\text{EQ}}$: At the start of this protocol, the parties have secret-shared values $[\![a]\!]$ and $[\![b]\!]$; at the end, they have a secret shared value $[\![c]\!]$, where $c = 1$ if $a = b$, and $c = 0$ otherwise.  

\item Secure less than test $\pi_{\text{LT}}$: At the start of this protocol, the parties have secret-shared values $[\![a]\!]$ and $[\![b]\!]$; at the end, they have a secret shared value $[\![c]\!]$, where $c = 1$ if $a < b$, and $c = 0$ otherwise.

\item Secure reveal $\pi_{\text{REVEAL}}$: At the start of this protocol, the parties have a secret-shared value $[\![a]\!]$, with the shares of $a$ spread across $K$ servers; at the end, the shares are aggregated and reconstructed so that the parties have the decrypted plaintext value $a$. 
\end{itemize}

We designed our MPC protocols using these cryptographic primitives to ensure that our final solution is also cryptographically secure, based on the universal composition theorem, which states that a protocol remains secure when composed with other or the same MPC protocols \cite{canetti2000security}.

For the experiments, we ran our MPC protocols in MP-SPDZ using both passive 3PC \cite{araki2016high} and active  3PC schemes \cite{eerikson2020use}. These MPC schemes execute over a power-of-two ring $\mathbb{Z}_{2^k}$, where we use the default ring size of $k = 64$ as in MP-SPDZ. For the computational costs of the primitives in these MPC schemes, please see the original publication of the MP-SPDZ library \cite{keller2020mp}.

To support fractional values, we employ fixed-point arithmetic (via the \texttt{sfix} type), where a real number $x$ is represented as a scaled integer with a default precision of $f = 16$ bits. By default, MP-SPDZ handles fixed-point truncation using binary-circuit-based decomposition. 
It works by decomposing the shared value into bits, performing the shift, and recomposing the values, providing exact truncation with no error. Because our multiplication products are bounded well within the 64-bit ring, this default protocol can operate correctly without any additional configuration.

Division by 2 and 6 in Protocol 3 is handled implicitly by MP-SPDZ's fixed-point arithmetic (\texttt{sfix}). Rather than performing division directly over $\mathbb{Z}_{2^{64}}$, MP-SPDZ converts division by a constant into multiplication by its fixed-point representation ($/2$ becomes $\times 0.5$ and $/6$ becomes $\times 0.1\overline{6}$, both representable as \texttt{sfix} values).




\newpage
\section{MPC Protocol for Irwin-Hall Approximation of Gaussian Noise}
\label{app:irwin_hall}
\begin{myprotocol}
\caption{$\pi_{\text{GAUSS}}$: MPC Protocol for Irwin-Hall Approximation of Gaussian Noise}
\label{alg:irwin_hall}
\footnotesize
\begin{algorithmic}[1]
\Require Target array size $Z$
\Ensure Secret shared array $[\![\vec{\mathcal{N}}]\!]$ of size $Z$ sampled from approximate $\mathcal{N}(0, 1)$

\State $[\![\vec{\mathcal{N}}]\!] \leftarrow \vec{0}$
\For{$k \leftarrow 1$ to 12} 
    \State $[\![\vec{\mathcal{N}}]\!] \leftarrow [\![\vec{\mathcal{N}}]\!] + \pi_{\text{RAND}}(0, 1, \text{size}=Z)$ \Comment{Sample from uniform distribution $\mathcal{U}(0,1)$}
\EndFor
\State $[\![\vec{\mathcal{N}}]\!] \leftarrow [\![\vec{\mathcal{N}}]\!] - \vec{6}$

\State \Return $[\![\vec{\mathcal{N}}]\!]$
\end{algorithmic}
\end{myprotocol}

Protocol \ref{alg:irwin_hall} for approximating standard Gaussian noise in secret-shared form by summing 12 independent uniform samples and subtracting 6, following the efficient Irwin–Hall method~\cite{pentyala2024caps}.

\newpage
\section{Additional Experimental Results}
\label{app:more_results}

\begin{table}[!ht]
  \centering
  \caption{\textbf{Utility.} Classification accuracy of a logistic regression model trained on synthetic data and evaluated on the real held-out test data across different numbers of genes, comparing 3PC active and 3PC passive schemes against centralized Private-PGM. All methods were evaluated at $\epsilon=10$.}
  \label{tab:accuracy_new}
  \setlength{\tabcolsep}{6pt}
  \begin{tabular}{llcccc}
    \toprule
    \textbf{{Protocol}} & \textbf{{Features}} & \textbf{{ALL}} & \textbf{{AML}} & \textbf{{BRCA}} & \textbf{{COMB}} \\
    \midrule
    \multirow{4}{*}{{Centralized}} & {200} & {$0.928$} & {$0.693$} & {$0.838$} & {$0.932$} \\
    & {500--600} & {$0.917$} & {$0.718$} & {$0.833$} & {$0.956$} \\
    & {800} & {$0.920$} & {$0.716$} & {$0.832$} & {$0.953$} \\
    & {959--1000} & {$0.916$} & {$0.705$} & {$0.830$} & {$0.958$} \\
    \midrule
    \multirow{4}{*}{{3PC Passive}} & {200} & {$0.934$} & {$0.697$} & {$0.827$} & {$0.939$} \\
    & {500--600} & {$0.918$} & {$0.717$} & {$0.826$} & {$0.952$} \\
    & {800} & {$0.921$} & {$0.706$} & {$0.832$} & {$0.956$} \\
    & {959--1000} & {$0.916$} & {$0.717$} & {$0.835$} & {$0.954$} \\
    \midrule
    \multirow{4}{*}{{3PC Active}} & {200} & {$0.941$} & {$0.699$} & {$0.838$} & {$0.937$} \\
    & {500--600} & {$0.920$} & {$0.727$} & {$0.833$} & {$0.951$} \\
    & {800} & {$0.918$} & {$0.702$} & {$0.833$} & {$0.952$} \\
    & {959--1000} & {$0.917$} & {$0.699$} & {$0.826$} & {$0.954$} \\
    \bottomrule
  \end{tabular}
\end{table}


{Table~\ref{tab:accuracy_new} presents the TSTR downstream utility for a LR model trained to classify the samples according to subtypes, measured on the held-out test set across a number of features (genes). For both 3PC passive and 3PC active MPC schemas, classification accuracy remains similar to the centralized baseline as the number of genes increases. Notably, the accuracy does not change significantly as the number of genes increases, showing that using a subset of genes is sometimes sufficient to achieve high accuracy for cancer subtype prediction.}

In Table \ref{tab:detpr_new}, we present the downstream biological value of the synthetic data by measuring the differential expression preservation (TPR of real differential expressed genes recovered in synthetic data) across a varying number of features (genes). Results for the centralized and federated 3PC approaches remain similar for all datasets, confirming that our MPC protocols preserve biological signals to the same extent as the centralized baseline. Across all protocols, DETPR decreases moderately as the feature count grows. For example, on BRCA, preservation drops from $\sim$0.77 at 200 features to $\sim$0.68 at 800, reflecting the increased difficulty of faithfully capturing marginal distributions as the number of genes increases while the privacy budget stays constant. Combined with the utility results in Tab.~\ref{tab:accuracy_new} and runtimes in Appendix~\ref{app:comm_cost}, this may be viewed as a tradeoff, where smaller feature sets yield stronger biological fidelity and faster MPC execution, but may sometimes lead to lower utility.


\begin{table}[!t]
  \centering
  \caption{{{\textbf{Biological value.} Differentially expressed gene preservation (DETPR; higher is better) for a varying number of genes, demonstrating the fraction of the original differentially expressed genes preserved in the synthetic data. All methods were evaluated at $\epsilon=10$.}}}
  \label{tab:detpr_new}
  \setlength{\tabcolsep}{6pt}
  \begin{tabular}{llcccc}
    \toprule
    \textbf{{Protocol}} & \textbf{{Features}} & \textbf{{ALL}} & \textbf{{AML}} & \textbf{{BRCA}} & \textbf{{COMB}} \\
    \midrule
    \multirow{4}{*}{{Centralized}} & {200} & {$0.567$} & {$0.582$} & {$0.778$} & {$0.924$} \\
    & {500--600} & {$0.519$} & {$0.556$} & {$0.706$} & {$0.893$} \\
    & {800} & {$0.511$} & {$0.543$} & {$0.680$} & {$0.869$} \\
    & {959--1000} & {$0.516$} & {$0.534$} & {$0.673$} & {$0.858$} \\
    \midrule
    \multirow{4}{*}{{3PC Passive}} & {200} & {$0.558$} & {$0.593$} & {$0.764$} & {$0.925$} \\
    & {500--600} & {$0.530$} & {$0.561$} & {$0.693$} & {$0.884$} \\
    & {800} & {$0.520$} & {$0.542$} & {$0.678$} & {$0.868$} \\
    & {959--1000} & {$0.504$} & {$0.533$} & {$0.669$} & {$0.856$} \\
    \midrule
    \multirow{4}{*}{{3PC Active}} & {200} & {$0.556$} & {$0.591$} & {$0.763$} & {$0.928$} \\
    & {500--600} & {$0.512$} & {$0.558$} & {$0.705$} & {$0.885$} \\
    & {800} & {$0.515$} & {$0.547$} & {$0.670$} & {$0.868$} \\
    & {959--1000} & {$0.519$} & {$0.535$} & {$0.663$} & {$0.858$} \\
    \bottomrule
  \end{tabular}
\end{table}



\begin{figure}[!h]
    \centering
    \begin{subfigure}[b]{0.45\textwidth}
        \centering
        \includegraphics[width=\textwidth]{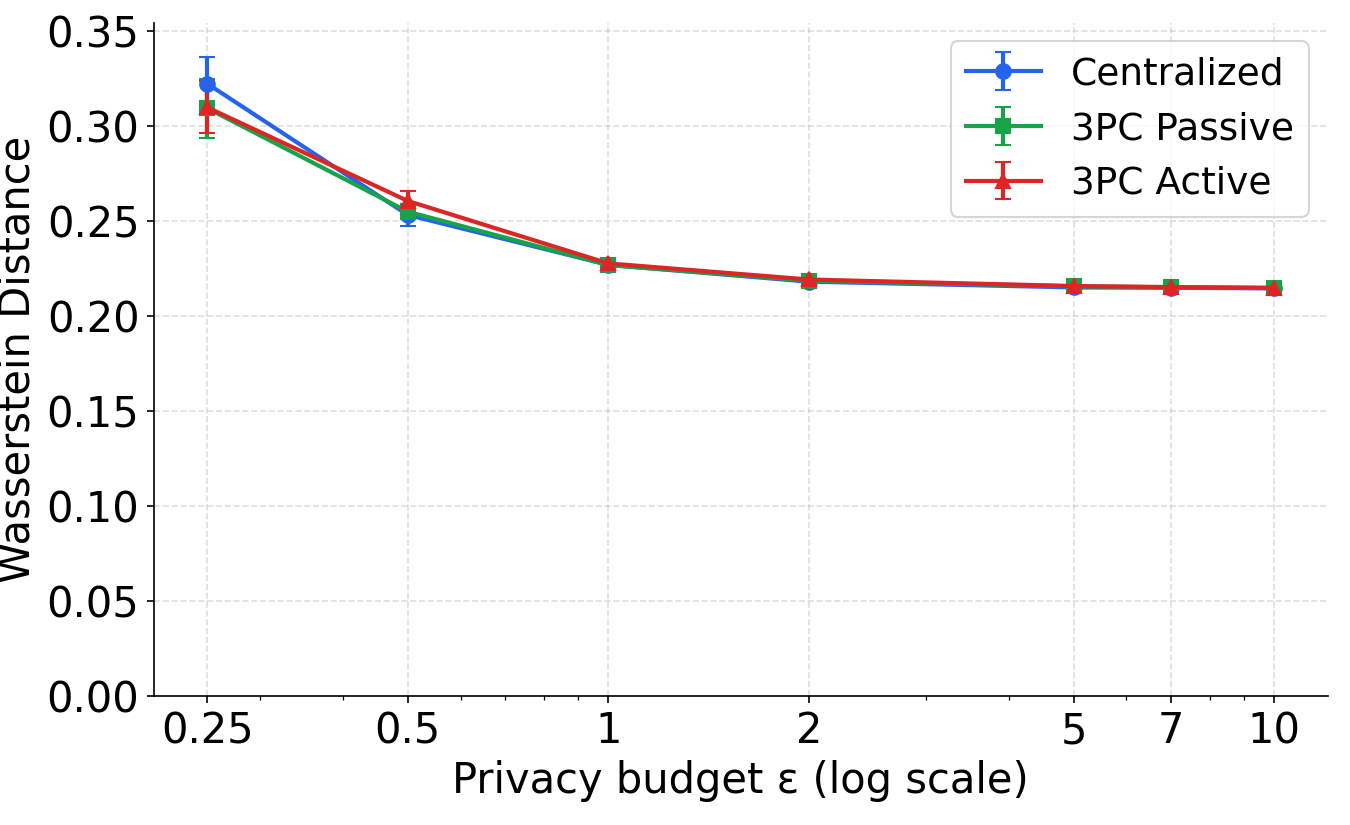}
        \caption{ALL}
        \label{fig:wasserstein_all}
    \end{subfigure}
    \hspace{0.02\textwidth}
    \begin{subfigure}[b]{0.45\textwidth}
        \centering
        \includegraphics[width=\textwidth]{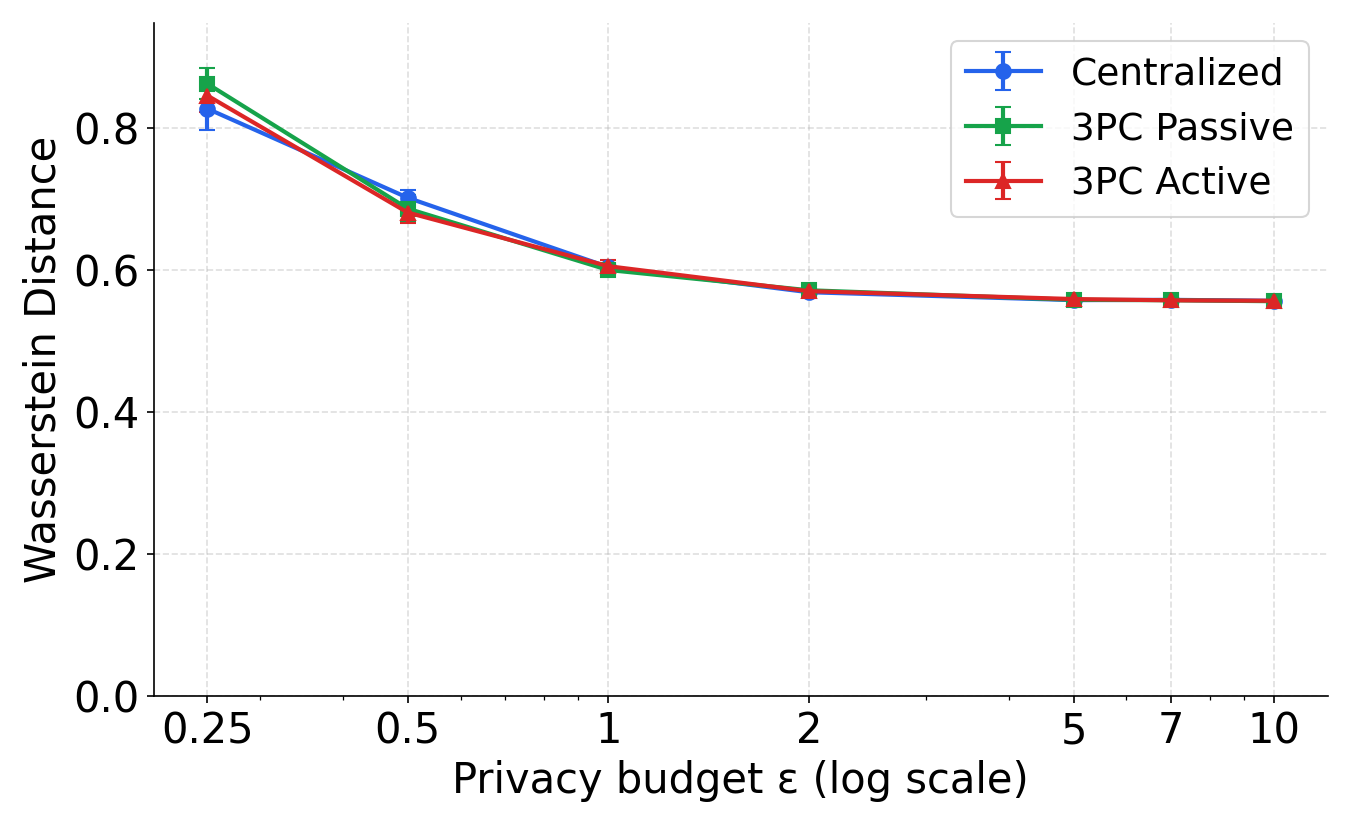}
        \caption{AML}
        \label{fig:wasserstein_aml}
    \end{subfigure}\\
    \begin{subfigure}[b]{0.45\textwidth}
        \centering
        \includegraphics[width=\textwidth]{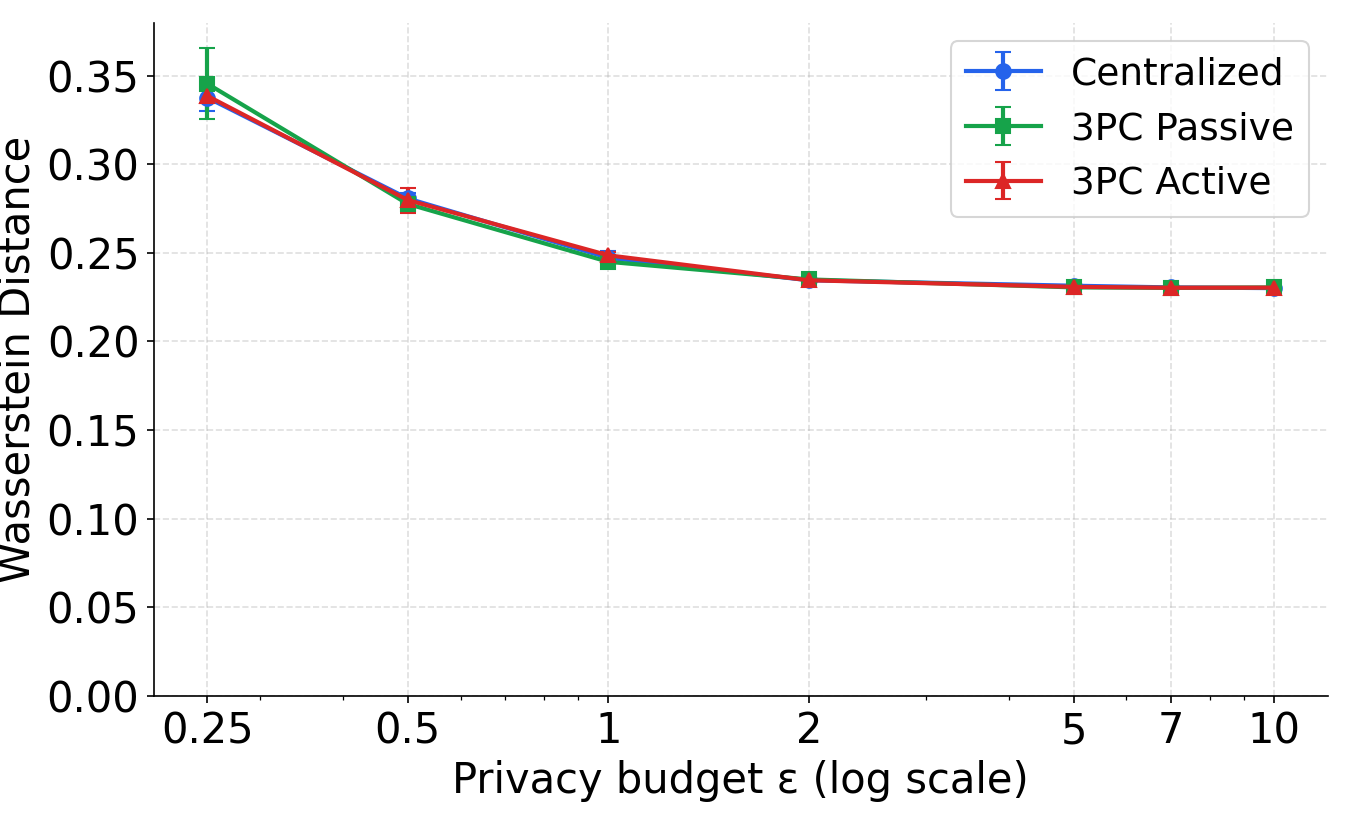}
        \caption{TCGA-BRCA}
        \label{fig:wasserstein_brca}
    \end{subfigure}
    \hspace{0.02\textwidth}
    \begin{subfigure}[b]{0.45\textwidth}
        \centering
        \includegraphics[width=\textwidth]{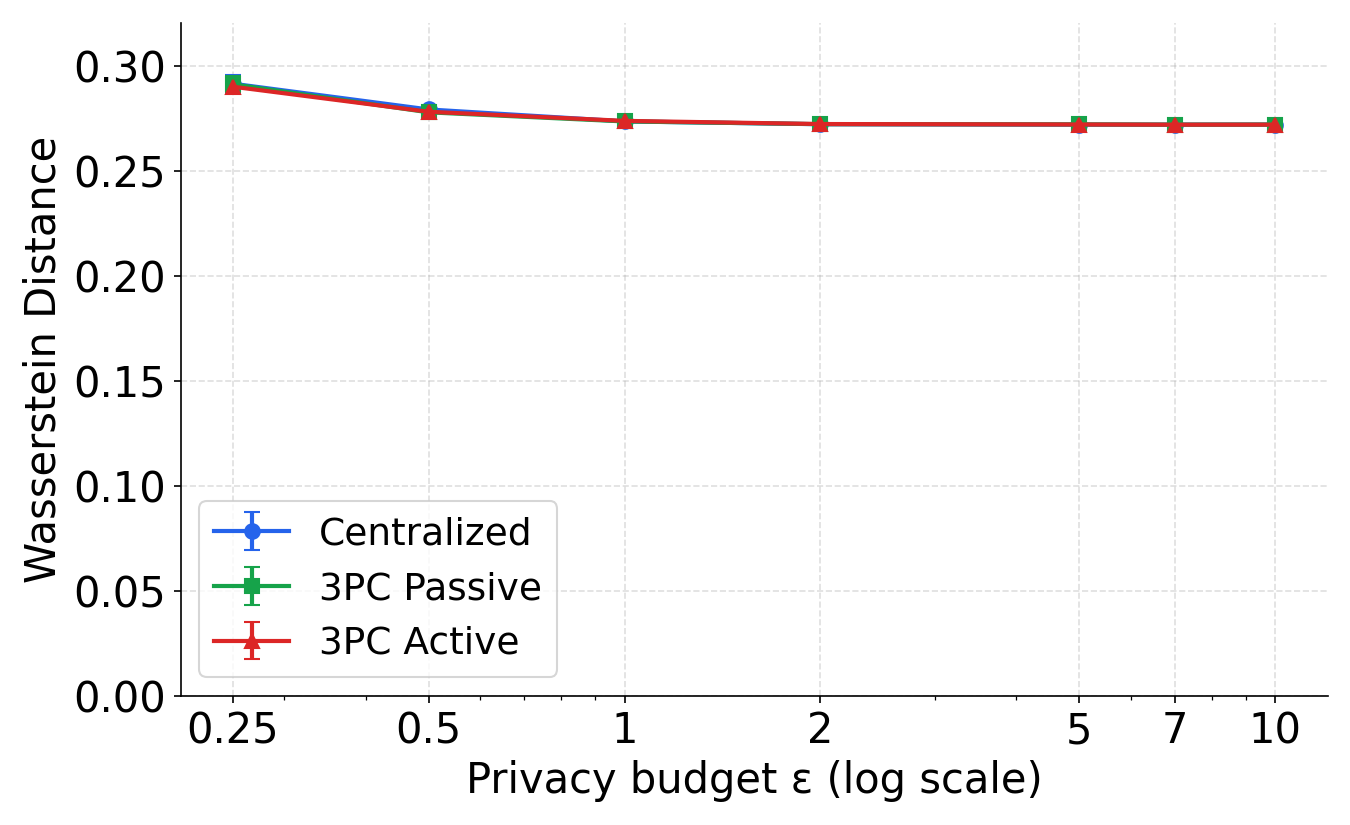}
        \caption{TCGA-COMBINED }
        \label{fig:wasserstein_comb}
    \end{subfigure}
    \caption{\textbf{Fidelity.} Wasserstein distance (lower is better) across privacy budgets across all datasets, demonstrating the distance between the original and the synthetic marginals.}
    \label{fig:wasserstein_scaling}
\end{figure}

\begin{figure}[!h]
    \centering
    \begin{subfigure}[b]{0.45\textwidth}
        \centering
        \includegraphics[width=\textwidth]{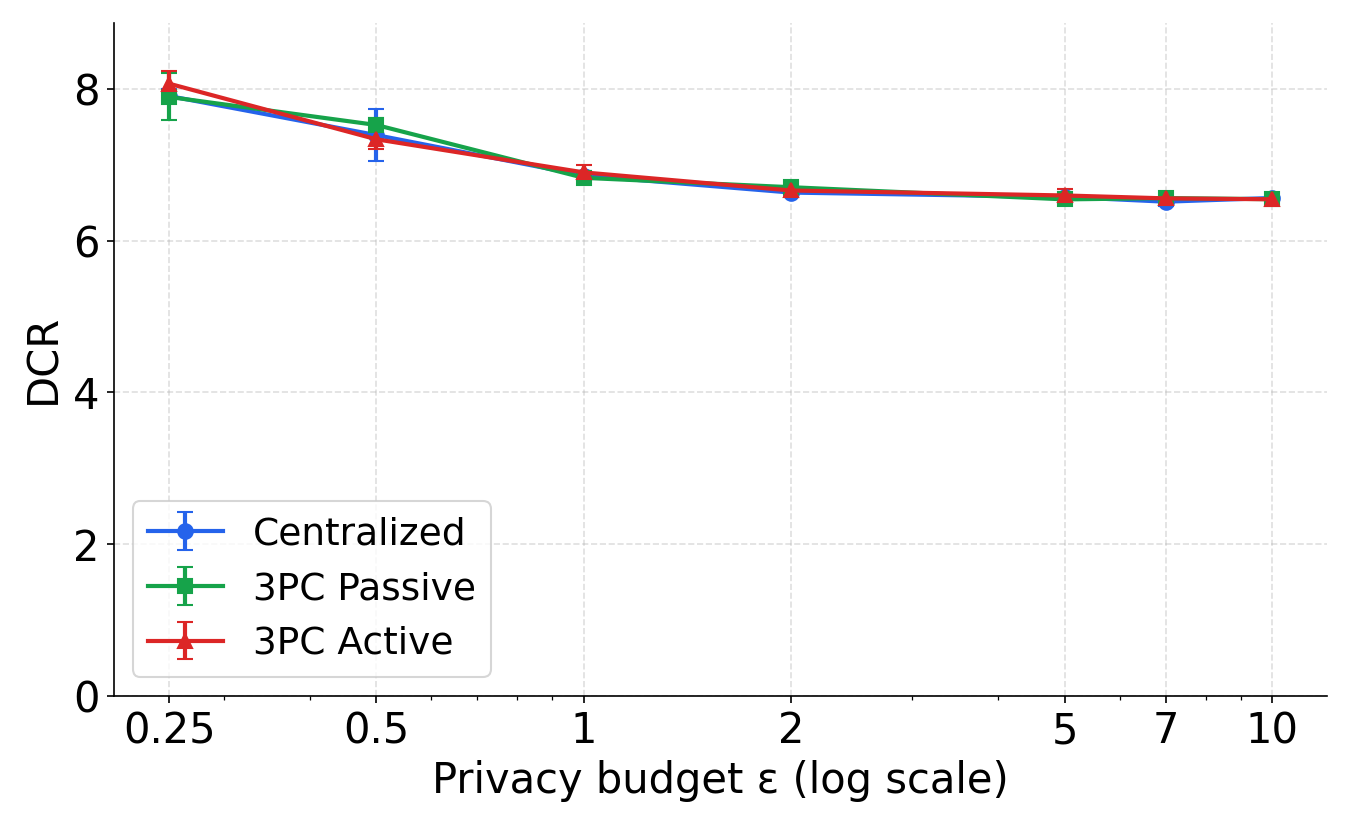}
        \caption{ALL}
        \label{fig:dcr_all}
    \end{subfigure}
    \hspace{0.02\textwidth}
    \begin{subfigure}[b]{0.45\textwidth}
        \centering
        \includegraphics[width=\textwidth]{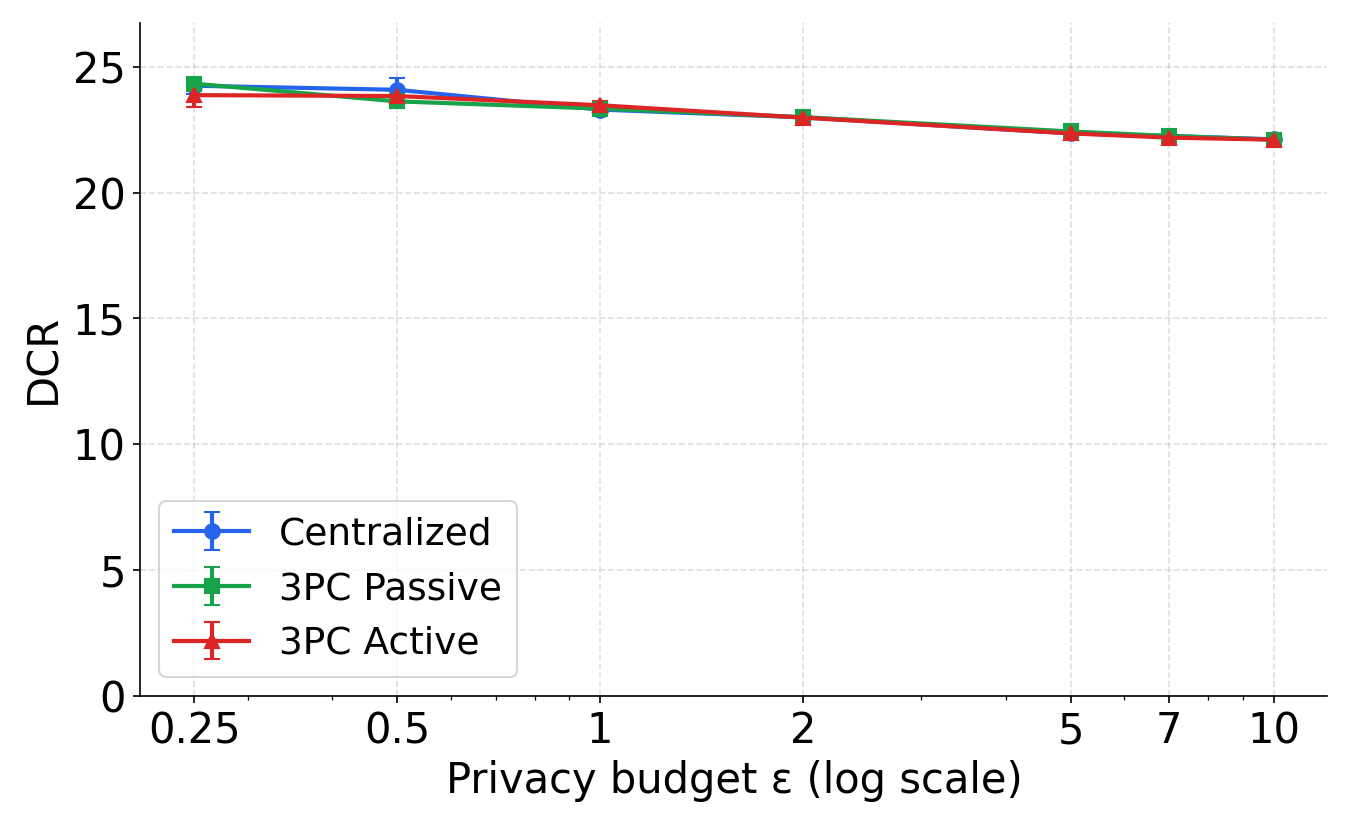}
        \caption{AML}
        \label{fig:dcr_aml}
    \end{subfigure}\\
    \begin{subfigure}[b]{0.45\textwidth}
        \centering
        \includegraphics[width=\textwidth]{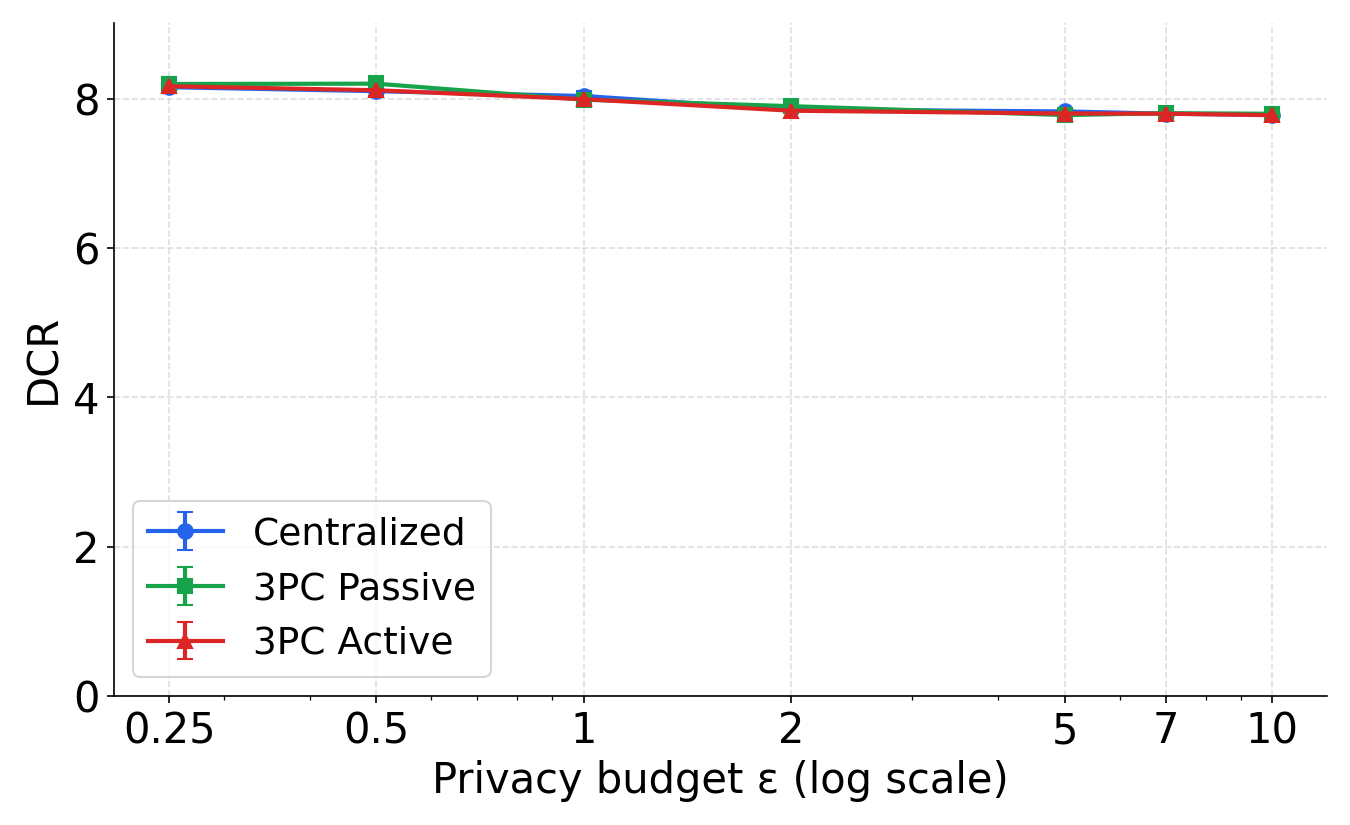}
        \caption{TCGA-BRCA}
        \label{fig:dcr_brca}
    \end{subfigure}
    \hspace{0.02\textwidth}
    \begin{subfigure}[b]{0.45\textwidth}
        \centering
        \includegraphics[width=\textwidth]{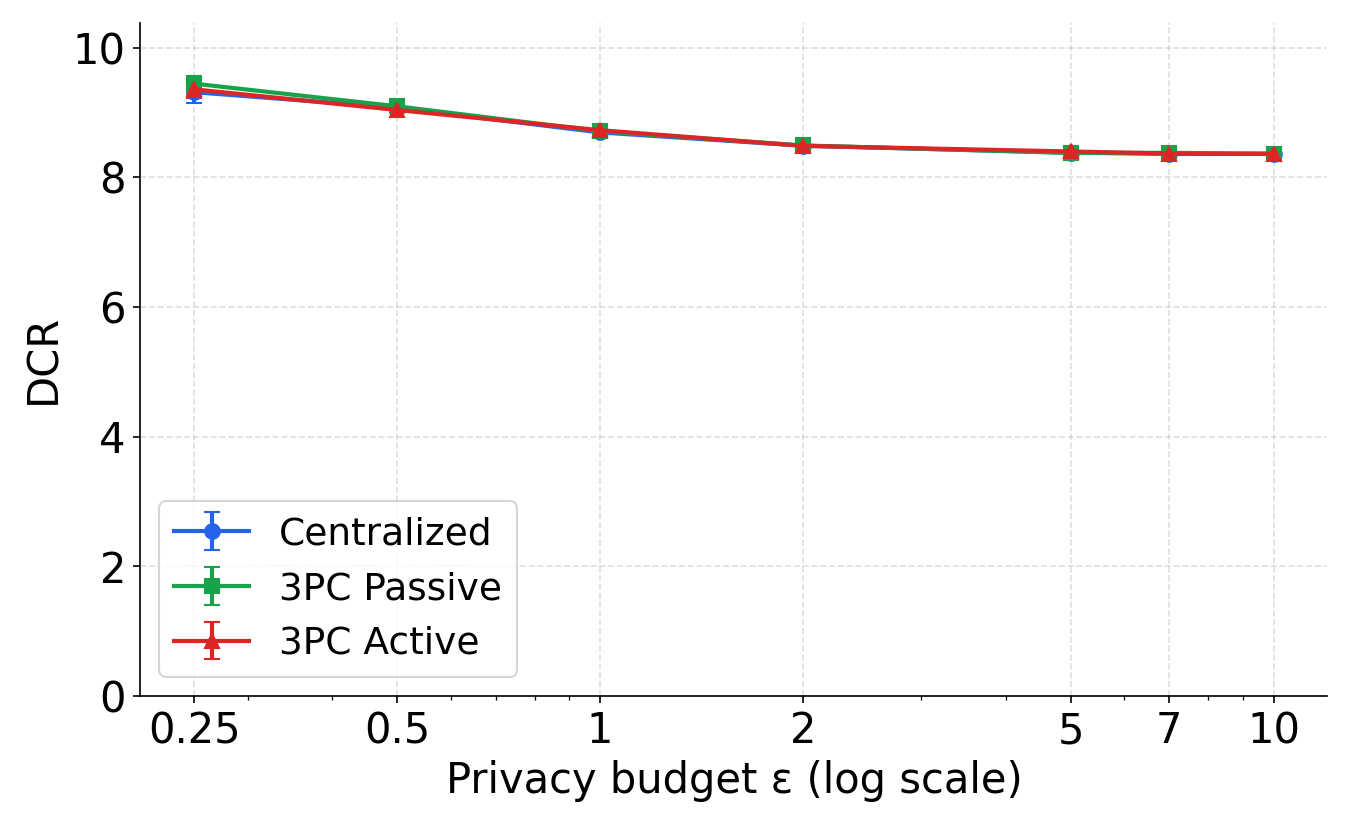}
        \caption{TCGA-COMBINED }
        \label{fig:dcr_comb}
    \end{subfigure}
    \caption{\textbf{Output privacy.} Distance to Closest Record (DCR; higher is better) across different privacy budgets $\epsilon$, acting as a proxy for memorization where a distance of 0 would entail that the generated synthetic data is identical to the real data and hence provides no privacy protection.} 
    \label{fig:dcr_privacy}
\end{figure}

\begin{figure}[!h]
    \centering
    \begin{subfigure}[b]{0.45\textwidth}
        \centering
        \includegraphics[width=\textwidth]{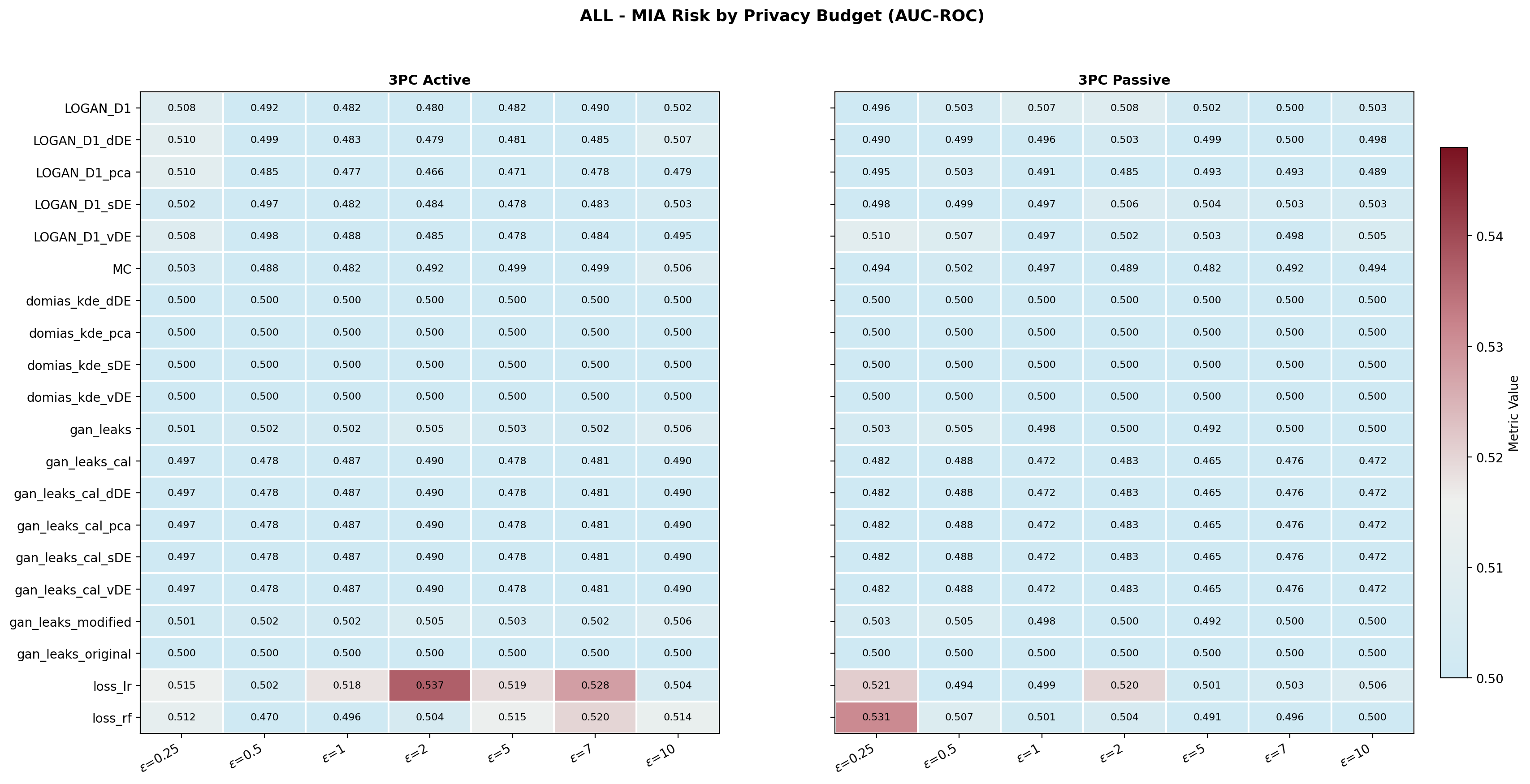}
        \caption{ALL}
        \label{fig:mia_all}
    \end{subfigure}
    \hspace{0.02\textwidth}
    \begin{subfigure}[b]{0.45\textwidth}
        \centering
        \includegraphics[width=\textwidth]{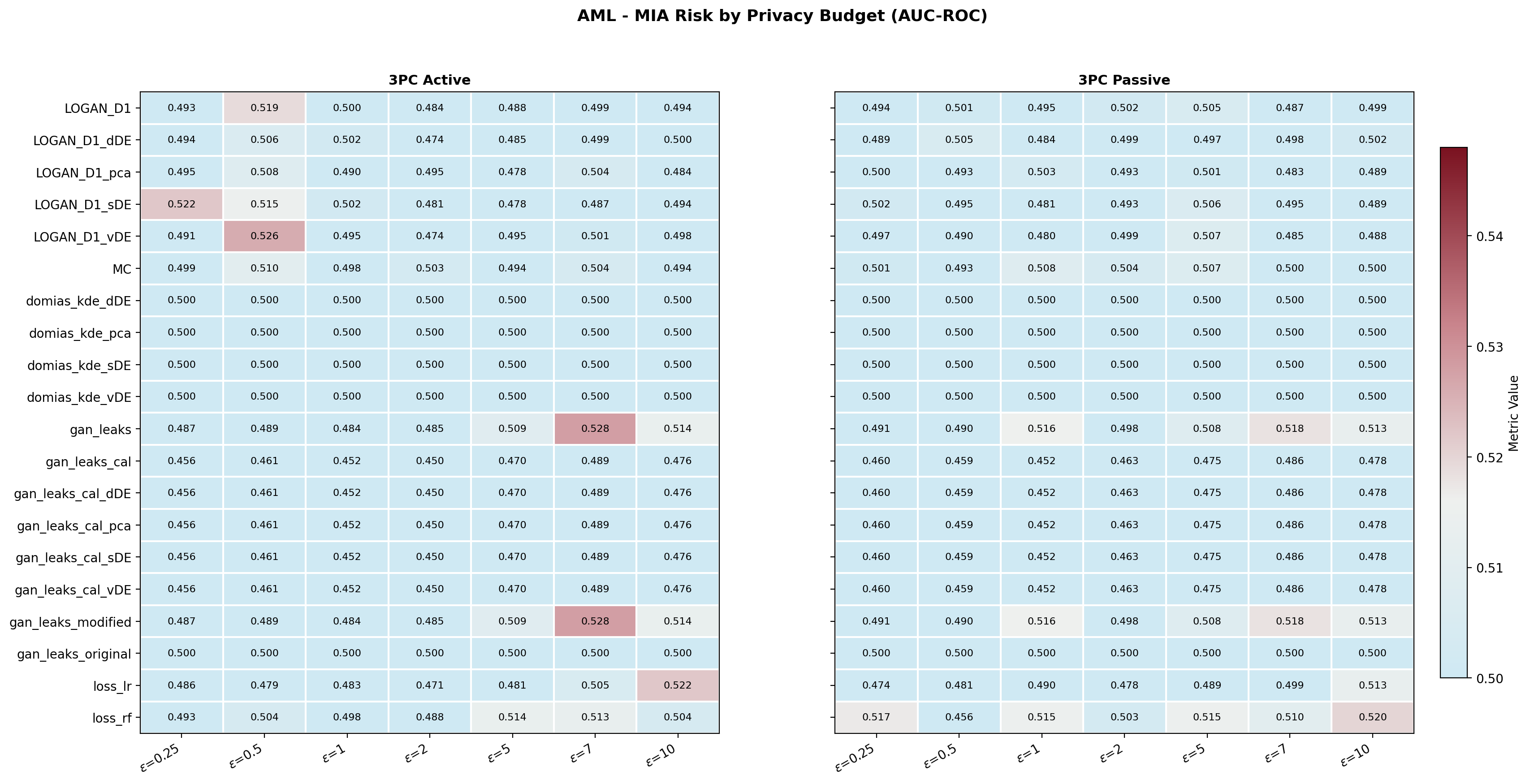}
        \caption{AML}
        \label{fig:dcr_aml}
    \end{subfigure}\\
    \begin{subfigure}[b]{0.45\textwidth}
        \centering
        \includegraphics[width=\textwidth]{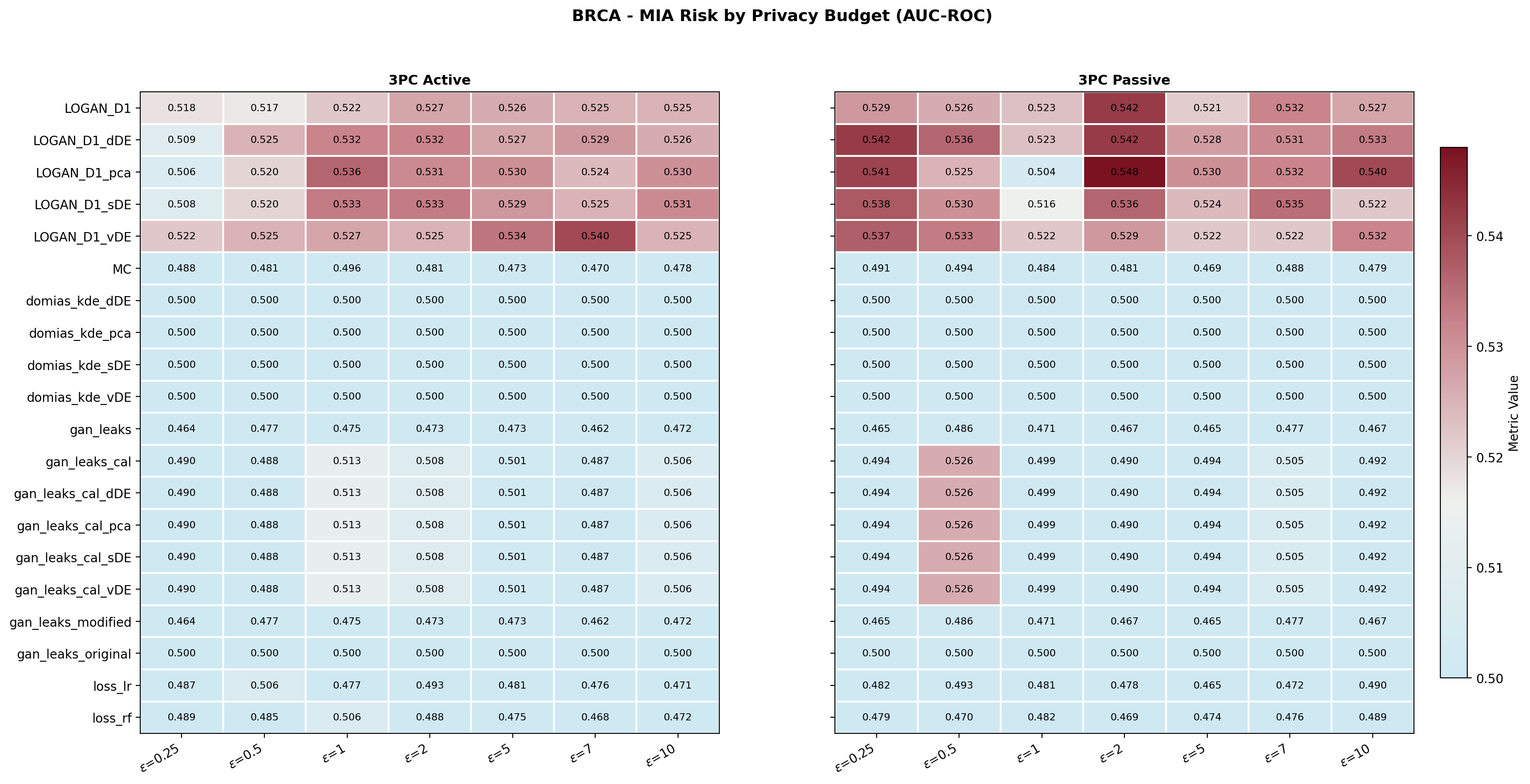}
        \caption{TCGA-BRCA}
        \label{fig:dcr_brca}
    \end{subfigure}
    \hspace{0.02\textwidth}
    \begin{subfigure}[b]{0.45\textwidth}
        \centering
        \includegraphics[width=\textwidth]{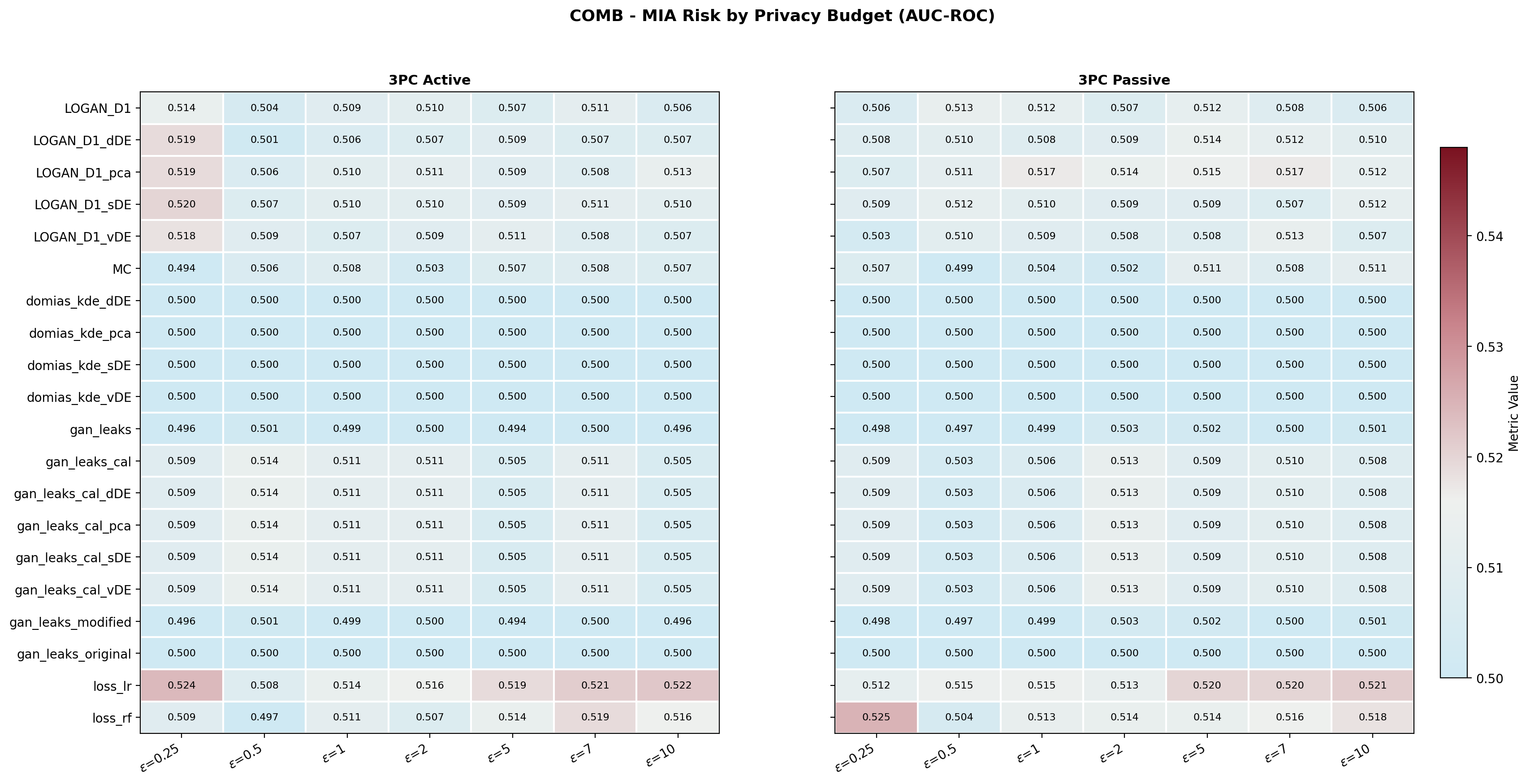}
        \caption{TCGA-COMBINED }
        \label{fig:dcr_comb}
    \end{subfigure}
    \caption{{\textcolor{black}{\textbf{Membership Inference Attacks.} (MIAs; AUC closer to 0.5 is better from a privacy perspective) across different privacy budgets $\epsilon$, representing the success of an adversarial attack that attempts to determine from synthetic data whether an instance was used to train the SDG model.}}}
    \label{fig:mia_privacy}
\end{figure}


Figure~\ref{fig:wasserstein_scaling} shows the fidelity of the synthetic marginals  for different privacy budgets, where a lower Wasserstein distance indicates better  quality. The results demonstrate that the proposed decentralized protocols, under both passive and active threat models, generate synthetic data with Wasserstein distances from the real data comparable to the centralized baseline, accurately reconstructing the underlying data distributions. As expected, tighter privacy budgets (smaller $\varepsilon$) lead to higher Wasserstein distances, reflecting the inherent privacy-fidelity trade-off: stronger privacy guarantees introduce more noise, which degrades distributional accuracy. This trend is consistent with the DCR results in Figure~\ref{fig:dcr_privacy}, where lower $\varepsilon$ similarly yields weaker record-level fidelity, confirming that the privacy-fidelity trade-off manifests coherently across both metrics.

{\textcolor{black}{Privacy of SDG methods is also commonly assessed in an empirical manner through Membership Inference Attacks (MIAs), i.e., adversarial attacks that aim to determine from the synthetic data whether a given instance was used as part of the training data for the generator \cite{golob2024privacy}. 
{Ozturk et al.~\cite{ozturk2026towards} report that Private-PGM in the centralized setting is resistant to state-of-the-art MIAs. The MIAs that they evaluate achieve an AUC around or only slightly above 0.5 (i.e., random guessing) on synthetic bulk RNAseq data generated by Private-PGM with $\epsilon=4$.
Figure~\ref{fig:mia_privacy} demonstrates that we arrive at similar findings in the 3PC passive and active settings, and across a range of $\epsilon$ values. The fact that state-of-the-art MIAs} (LOGAN \cite{hayes2019logan}, Monte-Carlo (MC) \cite{hilprecht2019monte}, DOMIAS \cite{vanbreugel2023membershipinferenceattackssynthetic}, GAN-leaks \cite{chen2020gan}) \textcolor{black}{operate near random guessing empirically verifies the privacy guarantees of our solution.}}}

\newpage
\section{Communication Costs and Per-Protocol Runtimes}\label{app:comm_cost}

Tables~\ref{tab:data_sent}--\ref{tab:protocol_runtimes} report the communication and
runtime overhead of our MPC protocols across all datasets and feature-set sizes.
In all three tables, \emph{3PC Passive} refers to the semi-honest setting \cite{araki2016high} and \emph{3PC Active} refers to the malicious setting \cite{eerikson2020use}. We observe that costs grow roughly linearly with the number of features, scaling with feature count.

\paragraph{Data sent (Tab.~\ref{tab:data_sent}).}
Tab.~\ref{tab:data_sent} reports the total volume of data (in GB) exchanged
among the three computing parties during a synthetic-data generation run.
The active setting incurs higher communication than the
passive setting across all datasets and feature counts, due to its greater complexity required to guarantee the resistance of the protocols to an active adversary.

\begin{table}[h]
\centering
\caption{Total data sent (GB) by dataset and number of features}
\label{tab:data_sent}
{\small
\begin{tabular}{llrrrr}
\toprule
\textbf{Dataset} & \textbf{MPC Scheme} & \multicolumn{4}{c}{\textbf{Number of Features}} \\
\cmidrule(lr){3-6}
 & & \textbf{200} & \textbf{500--600} & \textbf{800} & \textbf{959--1000} \\
\midrule
\multirow{2}{*}{ALL}
  & 3PC Passive &    3.61 &    9.03 &   14.44 &   17.31 \\
  & 3PC Active  &   83.18 &  207.92 &  332.67 &  398.78 \\
\midrule
\multirow{2}{*}{BRCA}
  & 3PC Passive &    3.34 &    8.33 &   13.33 &   16.30 \\
  & 3PC Active  &   81.60 &  203.97 &  326.35 &  398.96 \\
\midrule
\multirow{2}{*}{AML}
  & 3PC Passive &    3.80 &    9.48 &   15.17 &   18.95 \\
  & 3PC Active  &   84.65 &  211.55 &  338.45 &  423.05 \\
\bottomrule
\end{tabular}
}
\begin{tablenotes}
\footnotesize
\item ALL, BRCA, AML use feature counts 200/500/800/959--978--1000; COMB uses 200/600/800/979.
\item 3PC Passive = semi-honest (ring); 3PC Active = malicious (mal-rep-ring).
\end{tablenotes}
\end{table}

\bigskip
\begin{table}[!ht]
\centering
\caption{{\color{black}{Communication rounds by dataset and number of features}}}
\label{tab:comm_rounds}
{\small
{\color{black}
\begin{tabular}{lrrrr}
\toprule
\textbf{Dataset} & \multicolumn{4}{c}{\textbf{Number of Features}} \\
\cmidrule(lr){2-5}
 & \textbf{200} & \textbf{500--600} & \textbf{800} & \textbf{959--1000} \\
\midrule
ALL  &  53{,}189 & 132{,}943 & 212{,}697 & 254{,}966 \\
BRCA &  52{,}775 & 131{,}906 & 211{,}038 & 257{,}988 \\
AML  &  53{,}297 & 133{,}210 & 213{,}124 & 266{,}400 \\
COMB &  53{,}795 & 161{,}339 & 215{,}109 & 263{,}235 \\
\bottomrule
\end{tabular}
}
}
\begin{tablenotes}
\footnotesize
\item {\color{black}{Communication rounds are identical for 3PC Passive and 3PC Active since they share the same circuit.}}
\end{tablenotes}
\end{table}

\begin{table}[!ht]
\centering
\caption{{{MPC runtime per protocol organized by dataset and number of features}}}
\label{tab:protocol_runtimes}
{\small
{
\textbf{$\pi_{\text{BIN}}$}\\[4pt]
\begin{tabular}{llrrrr}
\toprule
\textbf{Dataset} & \textbf{MPC Scheme} & \multicolumn{4}{c}{\textbf{Number of Features}} \\
\cmidrule(lr){3-6}
 & & \textbf{200} & \textbf{500--600} & \textbf{800} & \textbf{959--1000} \\
\midrule
\multirow{2}{*}{ALL}
  & 3PC Passive &  25.3 s &  35.7 s &  51.8 s &   1.3 min \\
  & 3PC Active  & 14.9 min & 27.6 min & 45.2 min & 46.7 min \\
\midrule
\multirow{2}{*}{BRCA}
  & 3PC Passive &  15.7 s &  38.0 s &   1.0 min &   1.2 min \\
  & 3PC Active  &  9.7 min & 24.0 min & 39.1 min & 47.6 min \\
\midrule
\multirow{2}{*}{AML}
  & 3PC Passive &  15.8 s &  41.4 s &   1.1 min &   1.2 min \\
  & 3PC Active  & 10.0 min & 25.0 min & 50.2 min & 62.3 min \\
\midrule
\multirow{2}{*}{COMB}
  & 3PC Passive &  47.3 s &   2.4 min &   3.3 min &   3.8 min \\
  & 3PC Active  & 43.8 min & 119.9 min & 228.6 min & 267.9 min \\
\bottomrule
\end{tabular}
\vspace{8pt}

\textbf{$\pi_{\text{MARG}}$}\\[4pt]
\begin{tabular}{llrrrr}
\toprule
\textbf{Dataset} & \textbf{MPC Scheme} & \multicolumn{4}{c}{\textbf{Number of Features}} \\
\cmidrule(lr){3-6}
 & & \textbf{200} & \textbf{500--600} & \textbf{800} & \textbf{959--1000} \\
\midrule
\multirow{2}{*}{ALL}
  & 3PC Passive &  25.7 s &  29.4 s &  43.6 s &  55.3 s \\
  & 3PC Active  &  7.2 min & 20.9 min & 32.7 min & 33.6 min \\
\midrule
\multirow{2}{*}{BRCA}
  & 3PC Passive &  10.9 s &  27.2 s &  42.6 s &  51.8 s \\
  & 3PC Active  &  6.5 min & 16.2 min & 26.0 min & 31.6 min \\
\midrule
\multirow{2}{*}{AML}
  & 3PC Passive &  12.6 s &  30.2 s &   1.1 min &  59.3 s \\
  & 3PC Active  &  7.3 min & 18.3 min & 29.3 min & 37.2 min \\
\midrule
\multirow{2}{*}{COMB}
  & 3PC Passive &  39.5 s &   2.4 min &   3.3 min &   3.4 min \\
  & 3PC Active  & 25.8 min &  71.9 min & 107.8 min & 130.3 min \\
\bottomrule
\end{tabular}
\vspace{8pt}

\textbf{$\pi_{\text{BATCH-GAUSS}}$}\\[4pt]
\begin{tabular}{llrrrr}
\toprule
\textbf{Dataset} & \textbf{MPC Scheme} & \multicolumn{4}{c}{\textbf{Number of Features}} \\
\cmidrule(lr){3-6}
 & & \textbf{200} & \textbf{500--600} & \textbf{800} & \textbf{959--1000} \\
\midrule
\multirow{2}{*}{ALL}
  & 3PC Passive &   0.5 s &   0.6 s &   1.0 s &   1.2 s \\
  & 3PC Active  &   8.2 s &  23.2 s &  32.8 s &  39.0 s \\
\midrule
\multirow{2}{*}{BRCA}
  & 3PC Passive &   0.2 s &   0.6 s &   1.0 s &   1.1 s \\
  & 3PC Active  &   8.1 s &  20.3 s &  32.7 s &  39.8 s \\
\midrule
\multirow{2}{*}{AML}
  & 3PC Passive &   0.6 s &   1.5 s &   3.4 s &   3.1 s \\
  & 3PC Active  &  20.3 s &  51.2 s &   1.4 min &   1.7 min \\
\midrule
\multirow{2}{*}{COMB}
  & 3PC Passive &   0.4 s &   1.5 s &   2.3 s &   2.2 s \\
  & 3PC Active  &  14.9 s &  37.4 s &  59.9 s &   1.3 min \\
\bottomrule
\end{tabular}
}
}
\begin{tablenotes}
\footnotesize
\item ALL, BRCA, AML use feature counts 200/500/800/959--978--1000; COMB uses 200/600/800/979.
\item 3PC Passive = semi-honest (ring); 3PC Active = malicious (mal-rep-ring).
\end{tablenotes}
\end{table}

\textcolor{black}{
\paragraph{Communication Rounds (Tab.~\ref{tab:comm_rounds}).}
Tab.~\ref{tab:comm_rounds} lists the communication rounds recorded when running the protocol on different datasets, linearly scaling with the number of features in the dataset. The trend is consistent with the size of total data exchanged in the protocol (Tab.~\ref{tab:data_sent}). This is an important metric because communication is frequently a primary bottleneck for MPC protocols run over LAN/WAN.
}

\paragraph{Per-protocol runtimes (Tab.~\ref{tab:protocol_runtimes}).}
Tab.~\ref{tab:protocol_runtimes} breaks down the execution times for the three
sub-protocols introduced in Section~4: $\pi_{\text{BIN}}$, $\pi_{\text{MARG}}$,
and $\pi_{\text{BATCH-GAUSS}}$.
All three scale approximately linearly with the number of features.
$\pi_{\text{BATCH-GAUSS}}$ is the cheapest step, completing in under two minutes even in the active setting, followed by $\pi_{\text{MARG}}$ which requires an order of magnitude (anywhere from 20x to 100x) more time to execute. $\pi_{\text{BIN}}$ is the slowest protocol, consistently requiring over half of the total execution time to execute because of the expensive sorting operation. 

The active setting is consistently $30$--$60\times$ slower than the passive setting, reflecting the overhead of operations necessary to guarantee resistance against a malicious adversary.

The passive setting completes in a few minutes for all datasets, while the active setting requires up to roughly six hours for COMB at the largest feature count, a one-time offline cost per data release that does not affect downstream latency.


\newpage
\section{Comparison with Fu et al.}\label{app:comparisons}

{
Fu et al.~\cite{towardsfu2025} proposed a secure sort-and-count approach for marginal computation in the CaPS framework \cite{pentyala2024caps}. Their marginal computation protocol (equivalent to $\piMARG$ in Protocol \ref{alg:main_protocol}) is directly comparable to ours. Table~\ref{tab:runtime_comp} compares our $\piMARG$ against Fu et al.~across two datasets and varying feature counts. 
Our protocol is faster with gains increasing with dimensionality, i.e., number of features. Our approach is approximately 28\% faster on TCGA-BRCA and 32\% faster on TCGA-COMB. While the gap was recorded when running the protocols in 3PC Passive MPC schema, it will also exist in 3PC Active settings. This is consistent with the design intent that the vectorized dot-product formulation amortizes computation across features, so the advantage grows in the high-dimensional regime most relevant to RNA-seq applications.}




\begin{table}[!ht]
\centering
\caption{{Runtime comparison of 3PC Passive marginal computation protocol by Fu et al.~\cite{towardsfu2025} and our protocol. Time reported in seconds. Bold indicates the faster method.}}
\label{tab:runtime_comp}
\begin{tabular*}{\textwidth}{@{\extracolsep{\fill}}l l r r}
\toprule
{\textbf{Dataset}} & {\textbf{Features}} & {\textbf{Fu et al. Runtime (s)}} & {\textbf{Our Runtime (s)}} \\
\midrule
{\multirow{3}{*}{TCGA-BRCA (981 samples)}}
 & {100} & {5.79}  & {\textbf{5.58}} \\
 & {500} & {34.70} & {\textbf{26.48}} \\
 & {978} & {70.73} & {\textbf{50.84}} \\
\midrule
{\multirow{3}{*}{TCGA-COMB (3458 samples)}}
 & {100} & {27.13}  & {\textbf{21.14}} \\
 & {500} & {149.03} & {\textbf{103.07}} \\
 & {979} & {296.19} & {\textbf{200.50}} \\
\bottomrule
\end{tabular*}
\end{table}

{\color{black}{
\section{Hyperparameters and Implementation}
\label{app:hyperparameters}

\paragraph{Implementing DP in MPC.} Acknowledging the dangers of using DP with floating point arithmetic \cite{fu2026revisiting}, we follow the best practice of applying fixed-point and integer arithmetic as recommended by, for example, OpenDP \footnote{https://opendp.org/}. That finite precision may impact the exponential mechanism. If the utility of one of the classes in the exponential mechanism collapses to zero after the mapping into finite precision, pure DP becomes impossible to achieve. That can be dealt with by using approximate DP with the proper $\delta$ of magnitude inversely proportional to the size of the evaluation datasets. Such situation did not happen in our experiments.

Table~\ref{tab:hyperparameters} summarizes the MPC protocol and MP-SPDZ configuration used in all experiments. Unless otherwise stated,
the same configuration was used across datasets and privacy budgets.

\begin{table}[htbp]
\centering
\scriptsize
\caption{{\textcolor{black}{MPC protocol and model configuration.}}}
\label{tab:hyperparameters}
\color{black}
\begin{tabular}{p{0.45\linewidth} p{0.40\linewidth}}
\toprule
\textbf{Parameter} & \textbf{Value} \\
\midrule
Ring size & 64 \\
Data parties & 2 \\
Computing parties & 3 \\
Bins per feature & 4 \\
PGM inference iterations & 10,000 \\
\bottomrule
\end{tabular}
\end{table}

\end{document}